\documentstyle[aps,preprint]{revtex}
\begin{document}
{\sf
\title{
{\normalsize
\begin{flushright}
CU-TP949\\
RBRC-66
\end{flushright}}
Relations Between Low-lying Quantum Wave Functions and Solutions of 
the Hamilton-Jacobi Equation\thanks{Work supported in part by the US 
department of Energy}}

\author{
R. Friedberg$^{1}$, T. D. Lee$^{1-3}$ and W. Q. Zhao$^{2,4}$\\
{\small \it 1. Physics Department, Columbia University, New York, NY 10027, USA}\\
{\small \it 2. China Center of Advanced Science and Technology (CCAST)}\\
{\small \it         (World Lab.), P.O. Box 8730, Beijing 100080,  China}\\
{\small \it 3. RIKEN BNL Research Center (RBRC), Brookhaven National Lab.}\\
{\small \it         Bldg. 510, BNL, Upton, NY 11943, USA}\\
{\small \it 4. Institute of High Energy Physics, Academia Sinica, Beijing 
100039, China}  }
\maketitle

\newpage
\begin{abstract}
We discuss a new relation between the low lying Schroedinger wave function 
of a particle in a one-dimentional potential $V$ and the solution of the 
corresponding Hamilton-Jacobi equation with $-V$ as its potential. The 
function $V$ is $\geq 0$, and can have several minina ($V=0$). We assume the 
problem to be characterized by a small anhamornicity parameter $g^{-1}$ and 
a much smaller quantum tunneling parameter $\epsilon$ between these different
minima. Expanding  either the wave function or its energy as a formal double
power series in $g^{-1}$ and $\epsilon$, we show how the coefficients of 
$g^{-m}\epsilon^n$ in such an expansion can be expressed in terms of definite 
integrals, with leading order term determined by the classical solution of 
the Hamilton-Jacobi equation. A detailed analysis is given for the particular
example of quartic potential $V=\frac{1}{2}g^2(x^2-a^2)^2$.\\

\noindent
PACS{:~~11.10.Ef,~~03.65.Ge}
\end{abstract}

\newpage

\section*{\bf 1. Introduction}
\setcounter{section}{1}
\setcounter{equation}{0}

We discuss a hitherto unexplored link between the low-lying Schroedinger wave 
function of a particle in a potential $+V$ and the solution of the corresponding 
Hamilton-Jacobi equation with $-V$ as its potential. In this paper, we restrict 
our study to the one dimensional problem. Extension to multi-dimensional space will
be given in a separate publication.

Let $\Phi(x)$ be the solution of
\begin{eqnarray}\label{e1.1}
H\Phi(x) = E\Phi(x),
\end{eqnarray}
where 
\begin{eqnarray}\label{e1.2}
H = -\frac{1}{2}\frac{d^2}{dx^2} + V(x) 
\end{eqnarray}
describes the Hamiltonian of a non-relativistic particle of unit mass. Assume
\begin{eqnarray}\label{e1.3}
V(x) \geq 0;
\end{eqnarray}
however, its minimum $V(x) = 0 $ may occur at more than one point.

Introduce a scale factor $g^2$ by writing
\begin{eqnarray}\label{e1.4}
V(x) =g^2 v(x),
\end{eqnarray}
and consider the case of large $g$. We express 
\begin{eqnarray}\label{e1.5}
\Phi(x) = e^{-{\bf S}(x)}
\end{eqnarray}
in terms of a formal expansion:
\begin{eqnarray}\label{e1.6}
{\bf S}(x) = g{\bf S}_0(x) + {\bf S}_1(x) + g^{-1}{\bf S}_2(x) + g^{-2}{\bf S}_3(x) + \cdots
\end{eqnarray}
and likewise its eigenvalue
\begin{eqnarray}\label{e1.7}
E = gE_0 + E_1 + g^{-1}E_2 + g^{-2}E_3 + \cdots.
\end{eqnarray}
Substituting (\ref{e1.5}) - (\ref{e1.7}) into the Schroedinger equation (\ref{e1.1})
and equating the coefficients of $g^{-n}$ on both sides, we obtain
\begin{eqnarray}\label{e1.8}
{\bf S}_0'^2 &=& 2v, \nonumber\\
{\bf S}_0'{\bf S}_1' &=& \frac{1}{2} {\bf S}_0''-E_0,\\
{\bf S}_0'{\bf S}_2' &=& \frac{1}{2} ({\bf S}_1'' - {\bf S}_1'^2) -E_1 \nonumber
\end{eqnarray}
etc., where ${\bf S}_n' = d{\bf S}_n/dx$ and ${\bf S}_n'' = d^2{\bf S}_n/dx^2$. 
For each $n\geq 1$, the $(n+1)$th line of (\ref{e1.8})  determines ${\bf S}_n'$
uniquely in terms of the ${\bf S}_m$  for $m<n$. But the first line determines
${\bf S}_0'$  only  up  to an ambiguity in sign. By continuity, the ambiguity  
makes itself felt only at a global maximum of  $-v(x)$, that is where 
$v(x)$ vanishes quadratically and we must  decide whether  ${\bf S}_0'$
should behave analytically, changing  sign  as  it  passes through the
vanishing point, or whether ${\bf S}_0'$ should have a kink and take the same
sign on  both  sides  of  the  vanishing point.

If $v(x)$ vanishes only once, say at $x=a$, our choice is clear. ${\bf S}_0'$
should be positive for $x>a$ and negative for $x<a$, so  that ${\bf S}_0$ itself
will increase on  both sides  as  we go away from $x=a$, and  the 
wavefunction in  (\ref{e1.5}) will vanish at infinity in  both directions.
Thus  we should take  the  "analytic" square root in the first line of 
(\ref{e1.8}).

If there are $M>1$ values of $x$ where $v(x)$ vanishes, let us call one of  
them ($x=a$), where we want the wavefunction to  have its major peak, the 
{\it primary maximum} of $-v$, and  the  others {\it remote maxima}.  At the 
primary maximum ${\bf S}_0'$ should behave analytically  as  before.  But if we 
allow ${\bf S}_0'$ to change sign also at a remote maximum, we shall find ${\bf S}_0$ 
decreasing as we move farther on the other side, so that  the wavefunction 
may grow to another  peak competing  with the primary one, or may blow 
up at infinity. 

Thus, to have  a well-behaved wavefunction with no remote peaks, it appears 
that ${\bf S}_0'$ should change sign only at the primary  maximum,  and should
have a kink at the remote maxima.

Further support for  this  choice arises from the  generalization of our 
problem to many dimensions.  The first line of (\ref{e1.8}) will then  read
$(\bigtriangledown  {\bf S}_0)^2=2v$, and at the primary maximum it will
be natural to make $\bigtriangledown {\bf S}_0$ a vanishing vector with positive
divergence. Elsewhere ${\bf S}_0$ will be  found by integrating this 
(Hamilton-Jacobi) equation along a family of trajectories radiating out 
from the primary maximum. In general (excluding a set of measure zero)
these trajectories will not encounter any remote maximum. If we follow a
trajectory  that comes very near  to a remote maximum, we shall see $-v$
rising almost to zero but not quite. Thus, ${\bf S}_0'$ (differentiation 
along the trajectory) will not change sign, but will not quite have a 
kink.  Passing to the limit, we see that in order to make ${\bf S}_0$ 
continuous in all dimensions we should allow
${\bf S}_0'$ to develop a kink on the singular trajectory that passes through 
the remote maximum.

In the present paper we shall be concerned more with the quantum  mechanical 
wavefunction than with the Hamilton-Jacobi equation, as we wish to develop
corrections  to all orders in $g^{-1}$ and in $\epsilon$, the tunneling 
exponential. We  shall find it convenient to introduce two  types  of 
wavefunction denoted respectively by $\phi$ and $\psi$. Generically,
$\Phi$  refers to either $\phi$ or $\psi$.

The  groundstate energy will have an $M$-fold quasi-degeneracy  with a spread 
of  $O(\epsilon)$. Correspondingly, there are $M$ eigenstates  (called 
quasi-groundstates) of $H$. A $\phi$-function  will be an exact  solution of 
the  Schroedinger  equation  with an  energy  $E$  that is  {\it not} a true  
eigenvalue  but lies within the groundstate fine structure. A $\phi$-function
cannot  behave  well at  both  $+\infty$ and $-\infty$. On the other hand,
a $\psi$-function is a linear combination  of the true  quasi-groundstate
eigenfunctions. A $\psi$-function behaves well  at both infinities, but is  
not  an  exact solution of the Schroedinger equation  for any energy (except
when $\psi$ happens to be just one of the $M$ quasi-groundstate wave
functions).

In  this and the following section we shall study only the leading order
in $\epsilon$, and for this purpose we can restrict ourselves to the region
excluding any remote maximum, {\it i.e.}, to $x$ such that there is no second 
vanishing of $v$ between $x$ and $a$. In  this region $\phi$- and 
$\psi$-functions  are almost indistinguishable, and all the assertions in
these sections will apply indifferently to both, except for the exact
equations (\ref{e2.9}), (\ref{e2.10})  and (\ref{e2.14})  involving the
Wronskian.

In section 3 ff., we shall have to go beyond the remote 
maxima,  and  for  $M=2$  at least we  shall find that for  a $\phi$-function 
the  energy  $E$ can be  tuned so  that the  Hamilton-Jacobi solution   with
analytic ${\bf S}_0'$ remains  a  good approximation, whereas for a $\psi$-function  
the coefficients of the eigenfunctions can  be  tuned so that the
Hamilton-Jacobi solution with a kink in ${\bf S}_0'$  at the  remote maximum is
good.

It is  convenient to write the first line of (\ref{e1.8}) as
\begin{eqnarray}\label{e1.9}
\frac{1}{2}(\frac{d{\bf S}_0}{dx})^2 - v(x) = 0+, 
\end{eqnarray}
which denotes the corresponding Hamilton-Jacobi equation with $0+$ as the total
energy and $-v$ the potential. For its solution, we follow Hamilton's 
action-integral by starting at the primary maximum of $-v(x)$, say $x = a$; 
i.e.,
\begin{eqnarray}\label{e1.10}
-v(a) = 0.
\end{eqnarray}
At this starting point, the magnitude of ${\bf S}_0'$ is infinitesimal because the
total energy is $0+$. However, we will allow ${\bf S}_0'$ to change sign at $x=a$ in accordance with the discussion following (\ref{e1.8}).
Thus, ${\bf S}_0'$ is positive for $x>a$,
and negative for $x<a$. In the limit $0+\rightarrow 0$, ${\bf S}_0'=0$ at 
$x=a$ and also at any other (degenerate) maximum of $-v(x)$, say  $x=b$ and 
$b\not=a$. The limiting ${\bf S}_0'$ becomes analytic at $x=a$, but has a kink
at $x=b$. The resulting $\Phi=e^{-{\bf S}}$ describes  then a $\psi$-function.

Impose the ``normalization'' condition
\begin{eqnarray}\label{e1.11}
\Phi(a) = 1
\end{eqnarray}
and, correspondingly,
\begin{eqnarray}\label{e1.12}
{\bf S}_0(a) = {\bf S}_1(a) = {\bf S}_2(a) = \cdots = 0.
\end{eqnarray}
The solution of the Hamilton-Jacobi equation with the sign choices described
above is
\begin{eqnarray}\label{e1.13}
{\bf S}_0(x) = \pm \int\limits_a^x[2v(y)+(0+)]^{\frac{1}{2}}dy
\end{eqnarray}
with $[\cdots]^{\frac{1}{2}} $ denoting the principal value, the ``$+$'' sign for 
$x > a$ and the ``$-$'' sign for $x < a$, so that
\begin{eqnarray}
{\bf S}_0(x) \geq 0  ~~~~~~~~~~~~~~~~~{\sf everywhere.}\nonumber
\end{eqnarray}
As we shall see, the requirement that ${\bf S}_1(x),~{\bf S}_2(x),~\cdots$ be regular at $x=a$
determines $E_0,~E_1,~E_2,~\cdots$.

As a first illustration, consider the simple harmonic oscillator with
\begin{eqnarray}\label{e1.14}
V(x) = \frac{g^2}{2} x^2.
\end{eqnarray}
There is only one minimum of $V(x)$ at 
\begin{eqnarray}
x = a = 0. \nonumber
\end{eqnarray}
Thus, (\ref{e1.8}) and (\ref{e1.12}) - (\ref{e1.13}) lead to 
\begin{eqnarray}
{\bf S}_1(x) = {\bf S}_2(x) = \cdots = 0, \nonumber\\
E_1 = E_2 = \cdots =0, \nonumber
\end{eqnarray}
\begin{eqnarray}\label{e1.15}
{\bf S}(x) = g{\bf S}_0 = \frac{1}{2} g x^2
\end{eqnarray}
and
\begin{eqnarray}\label{e1.16}
E = g E_0 = \frac{1}{2} g.
\end{eqnarray}
The solution of the Hamilton-Jacobi equation
\begin{eqnarray}\label{e1.17}
\frac{1}{2}(\frac{d{\bf S}}{dx})^2 - \frac{g^2}{2}x^2 = 0+, 
\end{eqnarray}
gives the \underline{exact} groundstate Schroedinger wave function
\begin{eqnarray}\label{e1.18}
\Phi(x) = e^{-{\bf S}(x)} = e^{-\frac{1}{2}gx^2}. 
\end{eqnarray}
While the above approach resembles the familiar WKB method, it is different. The 
latter is good for excited levels, but not for the groundstate: the WKB wave 
function has an unnecessary ``classical turning point'' at 
$x = \pm g^{-\frac{1}{2}}$ (even though it does give the correct eigenvalue).
Note that WKB calls for introducing the energy $E$ on the right hand side of the
first line of (\ref{e1.8}), whereas we defer it to the second.

As a second example, we consider the quartic potential
\begin{eqnarray}\label{e1.19}
V(x) = \frac{1}{2}g^2 (x^2-a^2)^2.
\end{eqnarray}
The minimum $V(x) = 0$ is realized at two separate points
\begin{eqnarray}\label{e1.20}
x = \pm a.
\end{eqnarray}
In this problem, in addition to the small anharmonicity parameter
\begin{eqnarray}\label{e1.21}
(ga^3)^{-1} << 1,
\end{eqnarray}
there is a much smaller associated parameter
\begin{eqnarray}\label{e1.22}
\epsilon = e^{-\frac{4}{3}ga^3} 
\end{eqnarray}
that characterizes the tunneling between the two minima $x=a$ and $-a$. We shall 
start with the two solutions ${\bf S}_0(\pm)$ of the Hamilton-Jacobi equation 
(\ref{e1.9}),
\begin{eqnarray}\label{e1.23}
\frac{1}{2}[\frac{d}{dx}{\bf S}_0(\pm)]^2 - \frac{1}{2}(x^2 - a^2)^2 = 0+ 
\end{eqnarray}
by requiring
\begin{eqnarray}\label{e1.24}
&{\bf S}_0(+) = 0~~~~~~~~{\sf at}~~x = a \nonumber\\
{\sf and}&\\
&{\bf S}_0(-) = 0~~~~~~~~{\sf at}~~x = -a. \nonumber
\end{eqnarray}
Hence,
\begin{eqnarray}\label{e1.25}
&{\bf S}_0(+) = \pm \int\limits_a^x[(y^2-a^2)^2 + (0+)]^{\frac{1}{2}} dy \nonumber\\
{\sf and}&\\
&{\bf S}_0(-) = \pm \int\limits_{-a}^x[(y^2-a^2)^2 + (0+)]^{\frac{1}{2}} dy \nonumber
\end{eqnarray}
in accordance with (\ref{e1.13}) and the sign choices described below
(\ref{e1.10}), so that ${\bf S}_0(+)$ and ${\bf S}_0(-)$ are positive
everywhere. In addition, they are related to each other by 
\begin{eqnarray}\label{e1.26}
x \rightarrow -x, ~~~~~~~{\bf S}_0(\pm) \rightarrow {\bf S}_0(\mp).
\end{eqnarray}

Let H be the Hamiltonian associated with the quartic potential (\ref{e1.19}):
\begin{eqnarray}\label{e1.27}
H = -\frac{1}{2} \frac{d^2}{dx^2} + \frac{1}{2}g^2 (x^2-a^2)^2,
\end{eqnarray}
and $\psi_{even}$, $\psi_{odd}$ its even and odd ground-states with eigenvalues 
$E_{even}$ and $E_{odd}$. Write
\begin{eqnarray}\label{e1.28}
{\cal E} = \frac{1}{2} (E_{odd} + E_{even}), \nonumber\\
\Delta = \frac{1}{2} (E_{odd} - E_{even})
\end{eqnarray}
and
\begin{eqnarray}\label{e1.29}
\psi_{\pm}(x) = \frac{1}{2}[\psi_{even}(x) \pm \psi_{odd}(x)].
\end{eqnarray}
Thus,
\begin{eqnarray}\label{e1.30}
\psi_{+}(x) =\psi_-(-x)
\end{eqnarray}
\begin{eqnarray}\label{e1.31}
(H - {\cal E})\psi_{\pm}(x) & = &  -\Delta\psi_{\mp}(x)
\end{eqnarray}
and
\begin{eqnarray}\label{e1.32}
(\psi'_+\psi_- -\psi'_-\psi_+)' & = & 2\Delta \cdot(\psi^2_- -\psi^2_+),
\end{eqnarray}
where, as well as throughout the paper, the prime denotes $\frac{d}{dx}$.

We may expand ${\cal E}$ and $\Delta$ both formally as the following double series:
\begin{eqnarray}\label{e1.33}
{\cal E} = ga \sum\limits_{m,n} C_{mn}(ga^3)^{-m} e^{-\frac{4}{3}nga^3}
\end{eqnarray}
and
\begin{eqnarray}\label{e1.34}
\Delta = 4 (\frac{2}{\pi}g^3a^5)^{\frac{1}{2}} e^{-\frac{4}{3}ga^3}
       \sum\limits_{m,n} c_{mn}(ga^3)^{-m} e^{-\frac{4}{3}nga^3}
\end{eqnarray}
where $C_{mn}$ and $c_{mn}$ are numerical coefficients. The main body of this 
paper is to show how by using the solution (\ref{e1.25}) of the Hamilton-Jacobi 
Equation as the zeroth approximation to the Schroedinger wave function 
(\ref{e1.5}), we can obtain explicit expressions for these coefficients.

In Section 2, we derive
\begin{eqnarray}\label{e1.35}
C_{00} &=& c_{00} = 1, \nonumber\\
C_{10} &=& -\frac{1}{4}, ~~~~~ C_{20} = -9/2^6,\\
C_{30} &=& -89/2^9,~~~~{\sf etc.}\nonumber
\end{eqnarray}
The general formulas for $C_{mn}$ and $c_{mn}$ are obtained in Section 3 and 
4. Discussions for other examples of $V(x)$ are given in Section 5. In the 
Appendix, we analyze  further  the difference between a $\psi$-function 
and a $\phi$-function.

\newpage

\section*{\bf {2. Zeroth and First Orders in $\epsilon$}}
\setcounter{section}{2}
\setcounter{equation}{0}

We continue our analysis of the quartic potential case:
\begin{eqnarray}
V(x) = \frac{1}{2}g^2 (x^2-a^2)^2. \nonumber
\end{eqnarray}
To the zeroth order in
\begin{eqnarray}\label{e2.1}
\epsilon = e^{-\frac{4}{3}ga^3}, \nonumber\\
E_{even} \cong E_{odd} \cong E. 
\end{eqnarray}
We write in the notation of (\ref{e1.33}),
\begin{eqnarray}\label{e2.2}
E = ga \sum\limits_{m=0}^{\infty} C_{m0}(ga^3)^{-m}.
\end{eqnarray}
(This definition is a bit fuzzy since the series is asymptotic. In the next
section we shall define $E$ precisely. For the purpose of this section, it
suffices that $E \cong \frac{1}{2}(E_{even}+E_{odd})$ correct to
$O(\epsilon)$ and that (\ref{e2.2}) follows as a consequence. We define
$\phi_{\pm}$ as the exact solution of (\ref{e1.1}) with
$\phi_+(\infty)=\phi_-(-\infty)=0$. But since $E$ is not an eigenvalue of
$H$, $\phi_+(-\infty)$ and $\phi_-(\infty)$ are both divergent.)

For clarity, we reserve $\psi_{\pm}$ for the \underline{exact} wave functions given by (\ref{e1.29}), and use 
\begin{eqnarray}\label{e2.3}
\psi_+(x) \cong \phi_+(x)~~~~~~~{\sf for}~~~~~x>-a, \nonumber \\
\psi_-(x) \cong \phi_-(x)~~~~~~~{\sf for}~~~~~x<a
\end{eqnarray}
in the approximation of neglecting $\epsilon$.

Following (\ref{e1.5}) and setting the  generic $\Phi$ as $\phi_{\pm}$ we
write
\begin{eqnarray}\label{e2.4}
\phi_{\pm}(x) = e^{-S(\pm)}
\end{eqnarray}
with 
\begin{eqnarray}\label{e2.5}
&\phi_{+}(x) =\phi_-(-x), \nonumber\\
&\phi_{+}(\infty) =\phi_-(-\infty) = 0 \\
{\sf and}& \nonumber \\
&\phi_{+}(a) =\phi_-(-a) = 1. \nonumber
\end{eqnarray}

In accordance with (\ref{e1.6}) - (\ref{e1.7}) and writing  ${\bf S}$ as 
$S(\pm)$, we expand
\begin{eqnarray}\label{e2.6}
&S(\pm) = g \sum\limits_{m} g^{-m} S_m(\pm) \nonumber\\
{\sf and}&\\
&E = g \sum\limits_m g^{-m} E_m \nonumber
\end{eqnarray}
where 
\begin{eqnarray}\label{e2.7}
S_m(+) = 0~~~~~~~~{\sf at}~~x = a \nonumber\\
S_m(-) = 0~~~~~~~~{\sf at}~~x = -a,
\end{eqnarray}
and the corresponding $S_m'(+)$ and $S_m'(-)$ are finite at $x=a$ and $x=-a$,
respectively. As mentioned before (after (\ref{e2.2})), we have
\begin{eqnarray}\label{e2.8}
(H - E)\phi_{\pm} = 0,
\end{eqnarray}
which leads to
\begin{eqnarray}\label{e2.9}
(\phi'_+\phi_- -\phi'_-\phi_+)' = 0
\end{eqnarray}
and therefore
\begin{eqnarray}\label{e2.10}
\phi'_+\phi_- -\phi'_-\phi_+ = \lambda = const.
\end{eqnarray}
By using the Hamiltonian (\ref{e1.27}) and the set of equations (\ref{e1.8})
but replacing ${\bf S}_m$ by $S_m(\pm)$,  we derive

\newpage

\begin{tabular}{c|c}\hline
$|x+a| > O(\frac{1}{\sqrt{ga}})~~ {\sf for}~~S(+) $ &
$|x-a| > O(\frac{1}{\sqrt{ga}})~~ {\sf for}~~S(-) $ \\ \hline
$S_0(+) = \frac{1}{3}(x-a)^2(x+2a)$ & $S_0(-) = \frac{1}{3}(x+a)^2(-x+2a)$ \\ 
$S'_0(+) = x^2 - a^2$ & $S'_0(-) = -(x^2 - a^2)$ \\
$S_1(+) = ln\frac{x+a}{2a}$ & $S_1(-) = ln\frac{-x+a}{2a}$ \\  
$S'_1(+) = \frac{1}{x+a}$ & $S'_1(-) = \frac{1}{x-a}$ \\
$S_2(+) = \frac{3}{16a^3} - \frac{x + 2a}{4a^2(x + a)^2}$ & 
$S_2(-) = \frac{3}{16a^3} - \frac{-x + 2a}{4a^2(-x + a)^2}$ \\
$S'_2(+) = \frac{x + 3a}{4a^2(x+a)^3}$ & 
$S'_2(-) = -\frac{-x + 3a}{4a^2(-x+a)^3}$ \\
$S_3(+) =\frac{71}{2^9a^6} - \frac{9x^3 + 36ax^2 + 57a^2x + 40a^3}{64a^5(x+a)^4}$ & 
$S_3(-) =\frac{71}{2^9a^6} + \frac{9x^3 - 36ax^2 + 57a^2x - 40a^3}{64a^5(-x+a)^4}$\\
$S'_3(+) = \frac{9x^3 + 45ax^2 + 99a^2x + 103a^3}{64a^5(x+a)^5}$ & 
$S'_3(-) = \frac{9x^3 - 45ax^2 + 99a^2x - 103a^3}{64a^5(-x+a)^5}$\\ 
$S_4(+) = \frac{387}{2^{11}a^9}-$ &
$S_4(-) = S_4(+) (x \rightarrow -x)$\\
$-\frac{267x^5 +1602ax^4 +4094a^2x^3 +5766a^3x^2 +4779a^4x +2068a^5}
{3 \cdot 2^9a^8(x+a)^6}$&\\
$S'_4(+) = \frac{89x^5 + 623ax^4 + 1958a^2x^3 + 3594a^3x^2 + 4121a^4x + 2543a^5}
{512a^8(x + a)^7}$ &
$S'_4(-) = -S'_4(+) (x \rightarrow -x)$\\
\hline
\end{tabular}

\vspace{5mm}
\begin{eqnarray}\label{e2.11}
\end{eqnarray}

\begin{eqnarray}\label{e2.12}
E = ga -\frac{1}{4a^2} -\frac{9}{2g(2a)^5} -\frac{89}{2g^2(2a)^8} +
O(\frac{1}{g^3a^{11}})
\end{eqnarray}
and
\begin{eqnarray}\label{e2.13}
\lambda = 8ga^2e^{-\frac{4}{3}ga^3}[1 - \frac{3}{8ga^3} - 
\frac{53}{256(ga^3)^2} + \cdots].
\end{eqnarray}
This expression of $E$ gives the coefficients $C_{10}~-~C_{30}$ quoted in 
(\ref{e1.35}).

We  note that for $x<-2a$,  (\ref{e2.11}) gives a negative  $S_0(+)$, and as 
expected, when $x \rightarrow -\infty$, $S_0(+) \rightarrow -\infty$ and
therefore $\phi_+(-\infty)$ is divergent; likewise, as $x \rightarrow +\infty$, 
$S_0(-) \rightarrow -\infty$ and $\phi_-(\infty)$ is divergent.

In order to evaluate $\Delta$, we return to the exact wave functions
$\psi_{\pm}$, given by (\ref{e1.29}), and
integrate (\ref{e1.32}) from $-\infty$ to $x$.
Since $\psi_{even}(x)$ and $\psi_{odd}(x)$ are eigenfunctions of $H$, both are well
behaved at $x = \pm \infty$, so are $\psi_{\pm}(x)$ of (\ref{e1.29}). Hence, 
(\ref{e1.32}) gives
\begin{eqnarray}\label{e2.14}
\psi'_+(x)\psi_-(x) -\psi'_-(x)\psi_+(x) & = & 2\Delta \int\limits_{-\infty}^x
[\psi^2_-(y) -\psi^2_+(y)]dy.
\end{eqnarray}
According to (\ref{e2.3}) - (\ref{e2.4}) and (\ref{e2.6}), the zeroth order of
the exponents in $\psi_{\pm}=e^{-{\bf S}(\pm)}$ are determined by (\ref{e2.11})
provided $x>-a+ O(\frac{1}{\sqrt{ga}})$ for $\psi_+(x)$ and 
$x<a- O(\frac{1}{\sqrt{ga}})$ for $\psi_-(x)$. Take
\begin{eqnarray}\label{e2.15}
-a < x < 0
\end{eqnarray}
and also
$$ x+a >> (\frac{1}{ga})^{\frac{1}{2}}.$$
Hence, the left hand side of (\ref{e2.14}) is given by (\ref{e2.13}), while its 
right hand side is $2\Delta$ times
\begin{eqnarray}\label{e2.16}
\int\limits_{-\infty}^x [\psi^2_-(y) -\psi^2_+(y)]dy = \int\limits_{-\infty}^x
e^{-2S(-)}dy + O(\epsilon) \nonumber\\
~~~~~= \int\limits_{-\infty}^{\infty} e^{-2ga(y+a)^2}dy [1 +O(\frac{1}{ga^3}) +
O(\epsilon)] \nonumber \\
~~~~~=(\frac{\pi}{2ga})^{\frac{1}{2}} [1 +O(\frac{1}{ga^3}) +O(\epsilon)],
\end{eqnarray}
which leads to 
$$ 2\Delta \cong \lambda (\frac{2ga}{\pi})^{\frac{1}{2}},$$
and i.e.,
\begin{eqnarray}\label{e2.17}
\Delta = 4ga^2 (\frac{2ga}{\pi})^{\frac{1}{2}} e^{-\frac{4}{3}ga^3}
[1 +O(\frac{1}{ga^3}) +O(\epsilon)].
\end{eqnarray}
Thus, $c_{00} = 1$ in (\ref{e1.34}). By using the first line of (\ref{e2.16})
and expanding $exp[-S_1(-) -g^{-1}S_2(-) -g^{-2} S_3(-) \cdots]$ as a formal
series in $g^{-1}$, we can evaluate $c_{10},~c_{20},~\cdots$ in terms of
moments of Gaussian integrals.

In the next two sections, we develop a systematic method which enables us to 
evaluate all other $c_{mn}$ and $C_{mn}$ in (\ref{e1.33}) - (\ref{e1.34}) in 
terms of definite integrals.

\newpage

\section*{\bf 3. Definitions and Preliminaries }
\setcounter{section}{3}
\setcounter{equation}{0}

In this section we introduce several definitions and establish  some useful  
formulas for the quartic potential case.

In (\ref{e2.3}), the $\psi_{\pm}(x)$ on its left hand side are defined in 
terms of the eigenfunction $\psi_{even}$ and $\psi_{odd}$ of the 
quartic-potential  Hamiltonian $H$, in accordance with 
(\ref{e1.27}) - (\ref{e1.29}), whereas the $\phi_{\pm}(x)$ on its right 
hand side are introduced as solutions of 
\begin{eqnarray}\label{e3.1}
(H-E) \phi_{\pm}(x) = 0,
\end{eqnarray}
with $E$ between the two lowest eigenvalues $E_{even}$ and $E_{odd}$,
\begin{eqnarray}\label{e3.2}
E_{odd} > E > E_{even}, 
\end{eqnarray}
and $\phi_{\pm}$ satisfying the additional conditions
\begin{eqnarray}\label{e3.3}
&~~~~~~~~~~~~~~~~\phi_+(x) = \phi_-(-x) \nonumber\\
{\sf and}~~~~~~~~~~~&\\
&~~~~~~~~~~~~~~~~\phi_+(\infty) = \phi_-(-\infty) = 0. \nonumber
\end{eqnarray}
It follows from the Sturm-Liouville Theorem that $\phi_+(x)$ and $\phi_-(x)$ must 
have one (and only one) zero each at a finite $x$. Let 
\begin{eqnarray}\label{e3.4}
\phi_+(-\alpha) = \phi_-(\alpha) = 0.
\end{eqnarray}
When $E$ decreases from $E_{odd}$ to $E_{even}$, $\alpha$ moves from $x=0$ to 
$\infty$. It is convenient to choose $\alpha$ to  be near $a$, with
\begin{eqnarray}\label{e3.5}
|\alpha - a| \leq O(\frac{1}{\sqrt{ga}}).
\end{eqnarray}
As we shall see in Section 4, (\ref{e3.5}) is equivalent to constraint
\begin{eqnarray}\label{e3.6}
|E-\frac{1}{2}(E_{even} + E_{odd})| \leq O(\epsilon^2)
\end{eqnarray}
where $\epsilon$ is given by (\ref{e2.1}). Because of (\ref{e3.1}), 
$(\phi'_+\phi_- -\phi'_-\phi_+)'=0$ is exact, valid to
all orders of $\epsilon = e^{-\frac{4}{3}ga^3}$, and so is (\ref{e2.10}):
\begin{eqnarray}\label{e3.7}
\phi'_+\phi_- -\phi'_-\phi_+ = \lambda = const.
\end{eqnarray}
Likewise, the asymptotic expansion (\ref{e2.13})
\begin{eqnarray}
\lambda = 8ga^2e^{-\frac{4}{3}ga^3}[1 - \frac{3}{8ga^3} - 
\frac{53}{256(ga^3)^2} + \cdots] \nonumber
\end{eqnarray}
holds at all $x$. From (\ref{e3.3}) - (\ref{e3.4}) and (\ref{e3.7}), we have
\begin{eqnarray}\label{e3.8} 
\phi_+(x) = \lambda\phi_-(x)\cdot
\left\{\begin{array}{ccc}
\int \limits_{-\alpha}^x \frac{dy}{\phi^2_-(y)} & {\sf for} & x < \alpha  \\
-\int \limits_x^{\infty} \frac{dy}{\phi^2_-(y)} & {\sf for} & x > \alpha  
\end{array}
\right.
\end{eqnarray}
and
\begin{eqnarray}\label{e3.9} 
\phi_-(x) = \lambda\phi_+(x)\cdot
\left\{\begin{array}{ccc}
\int \limits_x^{\alpha} \frac{dy}{\phi^2_+(y)} & {\sf for} & x > -\alpha  \\
-\int \limits_{-\infty}^x \frac{dy}{\phi^2_+(y)} & {\sf for} & x < -\alpha.  
\end{array}
\right.
\end{eqnarray}
For $|x+a|>O(\frac{1}{\sqrt{ga}})$, the exponent $S(+)$ of
\begin{eqnarray}\label{e3.10}
\phi_+(x) = e^{-S(+)} 
\end{eqnarray}
has an asymptotic expansion given by (\ref{e2.6}) and (\ref{e2.11}). Through
\begin{eqnarray}\label{e3.11}
\phi_+(-x) = \phi_-(x) = e^{-S(-)},
\end{eqnarray}
we have the corresponding asymptotic expansion of $S(-)$ for 
$|x-a|>O(\frac{1}{\sqrt{ga}})$. These expressions and the Schroedinger 
equation  (\ref{e3.1}) lead to the previously derived asymptotic 
expansion (\ref{e2.12}) for $E$, independent of $\alpha$. 
To the zero$^{{\sf th}}$ order in $\epsilon$,
we note that for $\psi_+(x)$ of (\ref{e1.29}),  the asymptotic series is valid 
only for $x>-a+O(\frac{1}{\sqrt{ga}})$; however, as will be discussed later,
for $\phi_+(x)$ the same 
asymptotic series also holds for $x<-a-O(\frac{1}{\sqrt{ga}})$ if $\alpha$ 
satisfies (\ref{e3.5}). A similar difference exists between $\psi_-(x)$ 
and $\phi_-(x)$ through $x \rightarrow -x$.

Different choices of $\pm\alpha$ give different $\phi_{\pm}(x)$, and therefore 
also different $E$. The asymptotic expansion  for $E$ implies that these
different $E$ become the same to the zeroth order in $\epsilon$. (It will be
shown in Section 4 that actually the constraint (\ref{e3.5}) determines
$E$ to the first order in $\epsilon$.)

Similarly, for
$x>-a+O(\frac{1}{\sqrt{ga}})$, the $\alpha$-independent asymptotic expansion 
(\ref{e2.11}) for $S(+)$ implies that these different $\phi_+(x)$ are also the 
same if we neglect the $O(\epsilon)$ correction. For 
\begin{eqnarray}\label{e3.12}
x<-a+O(\frac{1}{\sqrt{ga}}), 
\end{eqnarray}
$\phi_+(x)$ is most conveniently given by
(\ref{e3.8}). We note that for  $x<a-O(\frac{1}{\sqrt{ga}})$, which includes the 
above region (\ref{e3.12}), the asymptotic expansion of (\ref{e2.11}) for $S(-)$
determines $\phi_-(x)$ to the zeroth order in $\epsilon$. Thus, in the region 
(\ref{e3.12}), under $\alpha \rightarrow \alpha + \delta \alpha$, we have from 
(\ref{e3.8})
\begin{eqnarray}\label{e3.13}
\delta\phi_+(x) = \lambda[\phi_-(x)/\phi_-^2(-\alpha)]\delta \alpha, 
\end{eqnarray}
in which $\lambda \delta \phi_- =  O(\epsilon^2)$ is neglected.  Since
$[1/\phi_-^2(y)]$ is minimum (for $y<\alpha$) at $y = -a$, the restriction 
(\ref{e3.5}) limits the corresponding $\delta \phi_+$  to $O(\epsilon)$  or
smaller. Likewise, through $x\rightarrow -x$, the corresponding $\delta\phi_-$
is also  $O(\epsilon)$ or smaller. Furthermore, for $\alpha$ satisfying the
restriction (\ref{e3.5}), in the region $x$ is less than $-a$ by 
$O(\frac{1}{\sqrt{ga}})$
\begin{eqnarray}
x<-a- O(\frac{1}{\sqrt{ga}}), \nonumber
\end{eqnarray} 
through the upper expression of (\ref{e3.8}) one can use the asymptotic 
expansion of $\phi_-=e^{-S(-)}$ to \underline{derive} the asymptotic expansion
of $\phi_+=e^{-S(+)}$, in accordance with (\ref{e2.11}). A similar statement 
holds in which $x \rightarrow -x$ and the roles of $\phi_-$ and $\phi_+$ are 
reversed.

In Section 4 we shall relate the eigenfunctions $\psi_{even}$ and $\psi_{odd}$
to these $\phi_{\pm}$. For this purpose it
is useful to introduce  two Green's  functions ("right" and "left"):
\begin{eqnarray}\label{e3.14}
<x|G_R|y> = \frac{2}{\lambda}\cdot
\left\{\begin{array}{cc}
-\phi_+(x)\phi_-(y) + \phi_-(x)\phi_+(y), & for\hspace{2mm} y > x \\
0 & for\hspace{2mm} x > y 
\end{array}
\right. 
\end{eqnarray} 
\begin{eqnarray}\label{e3.15}
<x|G_L|y> = \frac{2}{\lambda}\cdot
\left\{\begin{array}{cc}
0 & for\hspace{2mm} y > x  \\
\phi_+(x)\phi_-(y) - \phi_-(x)\phi_+(y), & for\hspace{2mm} x > y .  
\end{array} 
\right. 
\end{eqnarray} 
and two corresponding step functions: 
\begin{eqnarray}\label{e3.16}
S_R(x - y) = 
\left\{\begin{array}{cc}
-1 & if \hspace{2mm} y > x  \\
0 & if \hspace{2mm} x > y  
\end{array}
\right. 
\end{eqnarray} 
and 
\begin{eqnarray}\label{e3.17}
S_L(x - y) = 
\left\{\begin{array}{cc}
0 & if \hspace{2mm} y > x  \\
1 & if \hspace{2mm} x > y  
\end{array}
\right. 
\end{eqnarray} 
Hence, 
\begin{eqnarray}\label{e3.18}
<x|G_{R(L)}|y> = \frac{2}{\lambda}[\phi_+(x)\phi_-(y) - 
\phi_-(x)\phi_+(y)]\cdot S_{R(L)}(x-y).
\end{eqnarray} 
Both $S_R$ and $S_L$ satisfy 
\begin{eqnarray}\label{e3.19}
\frac {\partial}{\partial x}S_{R(L)}(x - y) = \delta(x - y). 
\end{eqnarray}
Since 
\begin{eqnarray}
[\phi_+(x)\phi_-(y) - \phi_-(x)\phi_+(y)]\delta (x-y) = 0 \nonumber 
\end{eqnarray}
we find
\begin{eqnarray}\label{e3.20}
\frac{\partial}{\partial x}<x|G_{R(L)}|y> & = &  
\frac{2}{\lambda}[\phi'_+(x)\phi_-(y) - 
\phi'_-(x)\phi_+(y)]S_{R(L)}(x-y),\\
-\frac{1}{2}\frac{\partial^2}{\partial x^2}<x|G_{R(L)}|y> & = &  
\frac{2}{\lambda}(-\frac{1}{2})[\phi''_+(x)\phi_-(y) - 
\phi''_-(x)\phi_+(y)]S_{R(L)}(x-y) \nonumber \\ 
& & + \frac{2}{\lambda}(-\frac{1}{2})[\phi'_+(x)\phi_-(y) - 
\phi'_-(x)\phi_+(y)]\delta (x-y) 
\end{eqnarray}
and 
\begin{eqnarray}\label{e3.22}
( H - E)<x|G_{R(L)}|y> = -\frac{1}{\lambda}
[\phi'_+(x)\phi_-(y) - \phi'_-(x)\phi_+(y)]\cdot \delta (x-y)
= - \delta(x - y) . 
\end{eqnarray}
Note that at a fixed $x$, as $y \rightarrow \infty$
\begin{eqnarray}\label{e3.23}
<x|G_{R}|y> \longrightarrow -\frac{2}{\lambda}\phi_+(x)\phi_-(y)
\longrightarrow \infty
\end{eqnarray}
and as $y\rightarrow -\infty$,
\begin{eqnarray}\label{e3.24}
<x|G_{L}|y> \longrightarrow -\frac{2}{\lambda}\phi_-(x)\phi_+(y)
\longrightarrow \infty
\end{eqnarray}
Because of (\ref{e3.7}) and (\ref{e3.10}) - (\ref{e3.11}), at all $x$ we have
\begin{eqnarray}\label{e3.25}
\phi_+(x)\phi_-(x) = \frac{\lambda}{S(-)'-S(+)'};
\end{eqnarray}
furthermore, for
\begin{eqnarray}\label{e3.26}
|x^2 - a^2| >> O(\sqrt{\frac{a}{g}}),
\end{eqnarray}
we find, by using (\ref{e2.11}),
\begin{eqnarray}\label{e3.27}
S(-)'-S(+)' = 2g(a^2-x^2) -\frac{2a}{a^2-x^2}-
\frac{3a^4+6a^2x^2-x^4}{2ga^2(a^2-x^2)^3}+O(\frac{1}{g^2}).
\end{eqnarray}
Therefore, integrals such as
\begin{eqnarray}\label{e3.28}
\int \limits_{-\infty}^{\infty}<x|G_{R}|y> \phi_+(y)dy 
\end{eqnarray}
and
\begin{eqnarray}\label{e3.29}
\int \limits_{-\infty}^{\infty}<x|G_{L}|y> \phi_-(y)dy
\end{eqnarray}
are convergent.

Next, we examine the equation
\begin{eqnarray}\label{e3.30}
H\phi_+ = -\frac{1}{2}\phi_+''+V\phi_+ = E \phi_+,
\end{eqnarray}
in which $E$ does not have to be an eigenvalue of $H$. Write, as before, 
$\phi_+ = e^{-S(+)}$ but introduce a new function $\theta$, defined to be
\begin{eqnarray}\label{e3.31}
\theta \equiv S(+)' - \sqrt{2V}.
\end{eqnarray}
Thus, $\theta$ satisfies
\begin{eqnarray}\label{e3.32}
\frac{1}{2}(\theta' - \theta^2) -\sqrt{2V} \theta +\frac{1}{2} \frac{V'}{\sqrt{2V}}
= E
\end{eqnarray}
which, for the quartic potential 
\begin{eqnarray}
V(x) = \frac{1}{2}g^2 (x^2-a^2)^2, \nonumber
\end{eqnarray}
becomes
\begin{eqnarray}\label{e3.33}
\frac{1}{2}(\theta' - \theta^2) -g(x^2-a^2)\theta +gx = E.
\end{eqnarray}
The boundary condition $\phi_+(\infty) = 0$  gives, for large $x$,
\begin{eqnarray}\label{e3.34}
\theta = \frac{1}{x} -\frac{E}{gx^2} +\frac{a^2}{x^3} +O(\frac{1}{gx^4}).
\end{eqnarray}
The asymptotic behavior of $\phi_+(x)$ as $x \rightarrow \infty$ is therefore 
insensitive to $E$; the same holds for $\phi_-(x)$ as $x \rightarrow -\infty$.
Consequently, the convergences of (\ref{e3.28}) - (\ref{e3.29}) also imply the
convergences of
\begin{eqnarray}\label{e3.35}
\int \limits_{-\infty}^{\infty}<x|G_{R(L)}|y> \psi_{even}(y)dy 
\end{eqnarray}
and
\begin{eqnarray}\label{e3.36}
\int \limits_{-\infty}^{\infty}<x|G_{R(L)}|y> \psi_{odd}(y)dy 
\end{eqnarray}

According to (\ref{e3.14}) - (\ref{e3.15}), each of these Green's functions $G_R$
and $G_L$ consists of two terms, $-\phi_+(x)\phi_-(y)$ and $\phi_-(x)\phi_+(y)$.
It is useful to examine their relative magnitudes. Since 
$\phi_+(-\alpha)= \phi_-(\alpha) = 0$, by using (\ref{e2.11}) we see that 
$\phi_+(x) > 0$ for $x>-\alpha$ and $\phi_-(x) > 0$ for $x<\alpha$. Because
\begin{eqnarray}\label{e3.37}
\big[\frac{\phi_-(x)}{\phi_+(x)}\big]' = - \frac{\lambda}{\phi_+^2(x)} < 0,
\end{eqnarray}
for any $x<y$ we have
\begin{eqnarray}\label{e3.38}
\frac{\phi_-(x)}{\phi_+(x)} > \frac{\phi_-(y)}{\phi_+(y)}.
\end{eqnarray}
Hence for $-\alpha<x<y$
\begin{eqnarray}\label{e3.39}
\phi_-(x)\phi_+(y) > \phi_-(y)\phi_+(x).
\end{eqnarray}
For 
\begin{eqnarray}\label{e3.40}
-\alpha < x < y < \alpha,
\end{eqnarray}
both sides of (\ref{e3.39}) are positive, which gives
\begin{eqnarray}\label{e3.41}
0 < \phi_-(x)\phi_+(y) -\phi_-(y)\phi_+(x) < \phi_-(x)\phi_+(y).
\end{eqnarray}
For
\begin{eqnarray}\label{e3.42}
\alpha < x < y,
\end{eqnarray}
both sides of (\ref{e3.39}) are negative. We have
\begin{eqnarray}\label{e3.43}
0 <-\phi_-(x)\phi_+(y) < -\phi_+(x)\phi_-(y)
\end{eqnarray}
and therefore
\begin{eqnarray}\label{e3.44}
0 < \phi_-(x)\phi_+(y) -\phi_+(x)\phi_-(y) <-\phi_+(x)\phi_-(y).
\end{eqnarray}
These inequalities help us to study the bounds on the integrals 
(\ref{e3.28}) - (\ref{e3.29}) and (\ref{e3.35}) - (\ref{e3.36}). For example, for
$x$ large $>>\alpha$, 
\begin{eqnarray}\label{e3.45}
\int \limits_{-\infty}^{\infty}<x|G_{R}|y> \phi_+(y)dy < -\frac{2}{\lambda}\phi_+(x)
\int\limits_x^{\infty}\phi_-(y)\phi_+(y)dy = -2\phi_+(x)\int\limits_x^{\infty}
\frac{1}{S(-)'-S(+)'}dy.
\end{eqnarray}
From (\ref{e3.27}) we find, as $x\rightarrow \infty$
\begin{eqnarray}\label{e3.46}
-2\int\limits_x^{\infty}\frac{1}{S(-)'-S(+)'}dy \longrightarrow \frac{1}{gx};
\end{eqnarray}
i.e.,
\begin{eqnarray}\label{e3.47}
\int \limits_{-\infty}^{\infty}<x|G_{R}|y> \phi_+(y)dy \longrightarrow \frac{1}{gx}
\phi_+(x).
\end{eqnarray}
By using (\ref{e3.31}) and (\ref{e3.34}), we see that as $x\rightarrow \infty$,
apart from a relative constant normalization factor between $\psi_{even(odd)}$
and $\phi_+$,
\begin{eqnarray}\label{e3.48}
\int \limits_{-\infty}^{\infty}<x|G_R|y> \psi_{even}(y)dy \longrightarrow 
\phi_+(x)/gx
\end{eqnarray}
and also
\begin{eqnarray}\label{e3.49}
\int \limits_{-\infty}^{\infty}<x|G_R|y> \psi_{odd}(y)dy \longrightarrow
\phi_+(x)/gx.
\end{eqnarray}

\newpage

\section*{\bf 4. Iterative Procedure }
\setcounter{section}{4}
\setcounter{equation}{0}

In this section, we present a systematic method to evaluate the eigenfunctions
$\psi_{even}$, $\psi_{odd}$ and the eigenvalues $E_{even}$ and $E_{odd}$ for the
quartic potential case (\ref{e1.27}). Write
\begin{eqnarray}\label{e4.1}
&~~~~~~~~~~~~E_{even} = E + \Delta_e \nonumber\\
{\sf and}~~~~~~~~~~~~~~&\\
&~~~~~~~~~~~~E_{odd} = E + \Delta_{od}, \nonumber
\end{eqnarray}
where $E$ satisfies (\ref{e3.2}) (and can be further tuned so that (\ref{e3.5})
holds).
By definitions, $\psi_{even}$ and $\psi_{odd}$ satisfy
\begin{eqnarray}\label{e4.2}
&~~~~~~~~~~~~(H-E)\psi_{even} = \Delta_e \psi_{even} \nonumber\\
{\sf and}~~~~~~~~~~~~~~&\\
&~~~~~~~~~~~~(H-E)\psi_{odd} = \Delta_{od} \psi_{odd}. \nonumber
\end{eqnarray}
From (\ref{e3.22}), we have
\begin{eqnarray}\label{e4.3}
( H - E) \cdot \Delta_j \int\limits_{-\infty}^{\infty}<x|G_R|y>\psi_j(y)dy = 
- \Delta_j \psi_j(x),
\end{eqnarray}
where $j=even(e)$ or $odd(od)$. Introduce
\begin{eqnarray}\label{e4.4}
\chi_j \equiv \psi_j +\Delta_j \int \limits_{-\infty}^{\infty}<x|G_R|y>
\psi_j(y)dy, 
\end{eqnarray}
which, on account of (\ref{e4.2}) - (\ref{e4.3}), satisfies
\begin{eqnarray}\label{e4.5}
(H-E)\chi_j(x) = 0.
\end{eqnarray}
This is the same as equation (\ref{e3.1}) for $\phi_+(x)$. Since both 
$\chi_j(x)$
and $\phi_+(x) \rightarrow 0$, as $x \rightarrow \infty$, we have
\begin{eqnarray}\label{e4.6}
\chi_j(x) \propto \phi_+(x).
\end{eqnarray}
From (\ref{e3.48}) - (\ref{e3.49}), the proportionality constant must not be 
zero, and can always be taken as unity by adjusting the normalization of 
$\psi_j$. Therefore, we can write
\begin{eqnarray}\label{e4.7}
\psi_{even}(x) = \phi_+(x) -\Delta_{e} \int \limits_{-\infty}^{\infty}
<x|G_R|y>\psi_{even}(y)dy 
\end{eqnarray}
and
\begin{eqnarray}\label{e4.8}
\psi_{odd}(x) = \phi_+(x) -\Delta_{od} \int \limits_{-\infty}^{\infty}
<x|G_R|y>\psi_{odd}(y)dy 
\end{eqnarray}
Impose the boundary conditions, at $x = 0$ 
\begin{eqnarray}\label{e4.9}
\psi'_{even}(0) = 0 \hspace{5mm}{\sf and}\hspace{5mm} \psi_{odd}(0) = 0 
\end{eqnarray}
From (\ref{e4.7}) - (\ref{e4.8}) and regarding $G_R$ as a square matrix, 
$\psi_{even}$, $\psi_{odd}$, $\phi_+$ as column matrices, we derive 
\begin{eqnarray}\label{e4.10}
\psi_{even(odd)} = (1 + \Delta_{e(od)}G_R)^{-1}\phi_+;
\end{eqnarray}
i.e.,
\begin{eqnarray}\label{e4.11}
\psi_{even} = (1 - \Delta_{e}G_R + \Delta_e^2G_R^2 - \cdots ) \phi_+ 
\end{eqnarray}
\begin{eqnarray}\label{e4.12}
\psi_{odd} = (1 - \Delta_{od}G_R + \Delta_{od}^2G_R^2 - \cdots ) \phi_+ 
\end{eqnarray}
Combining (\ref{e4.9}) with (\ref{e4.11}) and using $\phi'_+(0)=-\phi'_-(0)$, 
we find 
\begin{eqnarray}\label{e4.13}
1 + \frac{2\Delta_e}{\lambda}\int\limits_0^{\infty}
[\phi_+^2(y) + \phi_+(y)\phi_-(y)]dy &  &   \nonumber \\
- (\frac{2\Delta_e}{\lambda})^2\int\limits_0^{\infty}
[\phi_+(y_1) + \phi_-(y_1)]dy_1\int\limits_{y_1}^{\infty}
[\phi_-(y_1)\phi_+(y_2) - \phi_+(y_1)\phi_-(y_2)]
\phi_+(y_2)dy_2  & & \nonumber \\
+ \frac{2}{\lambda}\Delta_e^3\int\limits_0^{\infty}
[\phi_+(y_1) + \phi_-(y_1)]dy_1\int \limits_{-\infty}^{\infty} dy_2
 \int \limits_{-\infty}^{\infty} dy_3
<y_1|G_R|y_2><y_2|G_R|y_3>\phi_+(y_3)  & & \nonumber \\ 
+ \cdots &  = &  0. 
\end{eqnarray}
Likewise, combining (\ref{e4.9}) with (\ref{e4.12}), we derive
\begin{eqnarray}\label{e4.14}
1 - \frac{2\Delta_{od}}{\lambda}\int\limits_0^{\infty}
[\phi_+(y) - \phi_-(y)]\phi_+(y)dy  &  &  \nonumber \\
+ (\frac{2\Delta_{od}}{\lambda})^2\int\limits_0^{\infty}
[\phi_+(y_1) - \phi_-(y_1)]dy_1\int \limits_{y_1}^\infty
[\phi_-(y_1)\phi_+(y_2) - \phi_+(y_1)\phi_-(y_2)]\phi_+(y_2)dy_2 
& & \nonumber \\
- \frac{2}{\lambda}\Delta_{od}^3\int\limits_0^{\infty}
[\phi_+(y_1) - \phi_-(y_1)]dy_1\int \limits_{-\infty}^{\infty} dy_2
 \int \limits_{-\infty}^{\infty} dy_3
<y_1|G_R|y_2><y_2|G_R|y_3>\phi_+(y_3) & & \nonumber \\ 
+ \cdots & = & 0
\end{eqnarray}
From (\ref{e4.13}) - (\ref{e4.14}), $\Delta_e$ and $\Delta_{od}$ 
can be evaluated to successive orders of $\epsilon^n$. 

For a systematic analysis, we expand 
\begin{eqnarray}\label{e4.15}
\frac{2\Delta_{e}}{\lambda} = \delta_0(e) + \epsilon\delta_1(e) + 
\epsilon^2\delta_2(e) + \cdots + \epsilon^n\delta_n(e) + \cdots \nonumber \\ 
\frac{2\Delta_{od}}{\lambda} = \delta_0(od) + \epsilon\delta_1(od) + 
\epsilon^2\delta_2(od) + \cdots + \epsilon^n\delta_n(od) + \cdots 
\end{eqnarray}

Define 
\begin{eqnarray}\label{e4.16}
< x |f_R|y> & \equiv & \frac{\lambda}{2}<x|G_R|y> \nonumber \\
 & = & \left\{\begin{array}{cc}
\phi_-(x)\phi_+(y) - \phi_+(x)\phi_-(y), & x < y \\
0. & x > y. 
\end{array}
\right. 
\end{eqnarray}
Cast (\ref{e4.13}) and (\ref{e4.14}) into the forms 
\begin{eqnarray}\label{e4.17}
1 + a_0(\frac{2\Delta_e}{\lambda}) + \epsilon a_1(\frac{2\Delta_e}{\lambda})^2 
+ \epsilon^2 a_2(\frac{2\Delta_e}{\lambda})^3 + \cdots +  
\epsilon^n a_n(\frac{2\Delta_e}{\lambda})^{n+1} + \cdots = 0
\end{eqnarray}
and 
\begin{eqnarray}\label{e4.18}
1 + b_0(\frac{2\Delta_{od}}{\lambda}) + 
\epsilon b_1(\frac{2\Delta_{od}}{\lambda})^2 
+ \epsilon^2 b_2(\frac{2\Delta_{od}}{\lambda})^3 + \cdots +  
\epsilon^n b_n(\frac{2\Delta_{od}}{\lambda})^{n+1} + \cdots = 0
\end{eqnarray}
where
\begin{eqnarray}\label{e4.19}
a_0 = \int\limits_0^\infty [\phi_+^2(y) + \phi_-(y)\phi_+(y)]dy , 
\end{eqnarray} 
\begin{eqnarray}\label{e4.20}
b_0 = - \int\limits_0^\infty [\phi_+^2(y) - \phi_-(y)\phi_+(y)]dy , 
\end{eqnarray}  
\begin{eqnarray}\label{e4.21}
\epsilon a_1 = \int\limits_0^\infty [\phi_+(y_1) + \phi_-(y_1)]dy_1
\int \limits_{-\infty}^{\infty} <y_1|-f_R|y_2> \phi_+(y_2)dy_2, 
\end{eqnarray}   
\begin{eqnarray}\label{e4.22}
\epsilon b_1 = - \int\limits_0^\infty [\phi_+(y_1) - \phi_-(y_1)]dy_1
\int \limits_{-\infty}^{\infty} <y_1|-f_R|y_2> \phi_+(y_2)dy_2,  
\end{eqnarray}
\begin{eqnarray}\label{e4.23}
\epsilon^n a_n = \int\limits_0^\infty [\phi_+(y_1) + \phi_-(y_1)]dy_1
\int \limits_{-\infty}^{\infty} <y_1|(-f_R)^n|y_2> \phi_+(y_2)dy_2, 
\end{eqnarray}
\begin{eqnarray}\label{e4.24}
\epsilon^n b_n = - \int\limits_0^\infty [\phi_+(y_1) - \phi_-(y_1)]dy_1
\int \limits_{-\infty}^{\infty} <y_1|(-f_R)^n|y_2> \phi_+(y_2)dy_2.   
\end{eqnarray} 
\underline{Theorem}
\begin{eqnarray}\label{e4.25}
a_n = O(1) ~~~~~~{\sf and} ~~~~~~b_n = O(1). 
\end{eqnarray}
\underline{Proof}. Write
\begin{eqnarray}\label{e4.26}
\epsilon^n a_n = (-)^n J(+) ~~~~{\sf and} ~~~~
\epsilon^n b_n = -(-)^n J(-),
\end{eqnarray}
where, in accordance with (\ref{e4.23}) - (\ref{e4.24}),
\begin{eqnarray}\label{e4.27}
J(\pm) \equiv \int \limits^\infty_0 dy_1\int \limits^\infty_{y_1} dy_2 
\cdots \int \limits^\infty_{y_{n-1}} dy_n\int \limits^\infty_{y_n} dy_{n+1} I(\pm)
\end{eqnarray}
and 
\begin{eqnarray}\label{e4.28}
I(\pm) & = & [\phi_+(y_1) \pm \phi_-(y_1)]\cdot[\phi_-(y_1)\phi_+(y_2) - 
\phi_+(y_1)\phi_-(y_2)] \cdots\nonumber \\
& & \cdot [\phi_-(y_m)\phi_+(y_{m+1}) - 
\phi_+(y_{m})\phi_-(y_{m+1})]\cdots \nonumber \\
& & \cdot[\phi_-(y_n)\phi_+(y_{n+1}) - 
\phi_+(y_{n})\phi_-(y_{n+1})]\cdot\phi_+(y_{n+1}).
\end{eqnarray}
Decompose the integration range of (\ref{e4.27}) into the sum  
$$(0) + (1) + (2) + \cdots + (m) + \cdots (n) + (n+1), $$
where
\begin{eqnarray}\label{e4.29}
(0): \hspace{5mm}\alpha < y_1 < y_2 < \cdots < y_n < y_{n+1} 
\end{eqnarray}
\begin{eqnarray}\label{e4.30}
(m): &\hspace{5mm}0< y_1 < y_2< \cdots < y_m < \alpha, \nonumber\\
&\hspace{5mm} \alpha < y_{m+1} < \cdots < y_n < y_{n+1} 
\end{eqnarray}
with $m=1, 2, \cdots, n$, and
\begin{eqnarray}\label{e4.31}
(n+1):\hspace{5mm} 0< y_1 < y_2< \cdots < y_n < y_{n+1} < \alpha 
\hspace{5mm}.
\end{eqnarray} 
The integral (\ref{e4.27}) can be written as 
\begin{eqnarray}\label{e4.32}
J(\pm) = \sum\limits^{n+1}_{m=0}J_m(\pm),
\end{eqnarray} 
where  
\begin{eqnarray}\label{e4.33}
J_m(\pm) = \int_{(m)}\prod\limits_1^{n+1}dy_i I(\pm) = 
\int\limits^\alpha_0 dy_1\int\limits^\alpha_{y_1} dy_2 \cdots 
\int\limits^\alpha_{y_{m-1}} dy_{m}\int\limits_\alpha^\infty dy_{m+1}
\int\limits_{y_{m+1}}^\infty dy_{m+2}\cdots\int\limits_{y_n}^\infty 
dy_{n+1}\cdot I(\pm).
\end{eqnarray} 

Consider first $J_0$. From (\ref{e3.44}) and (\ref{e4.28}), the magnitude of 
$I(\pm)$ satisfies the inequality
\begin{eqnarray}\label{e4.34}
|I(\pm)|_{in(0)} & < & (-)^n |\phi_+(y_1) \pm \phi_-(y_1)|
[\phi_+(y_1)\phi_-(y_2)][\phi_+(y_2)\phi_-(y_3)]\cdots \nonumber \\
& & [\phi_+(y_n)\phi_-(y_{n+1})]\phi_+(y_{n+1}).
\end{eqnarray}
Since at any $x$, on account of (\ref{e3.25}),
\begin{eqnarray}\label{e4.35}
\phi_+(x)\phi_-(x) = O(\epsilon) 
\end{eqnarray}
and for $x>-\alpha$, 
\begin{eqnarray}\label{e4.36}
\phi_+^2(x) \leq O(1),
\end{eqnarray}
we have 
\begin{eqnarray}\label{e4.37}
|\phi_+(y_1) \pm \phi_-(y_1)|\phi_+(y_1) \leq O(1)
\end{eqnarray}
and therefore
\begin{eqnarray}\label{e4.38}
J_{m=0}(\pm) = O(\epsilon^n).
\end{eqnarray}
Next, for any $m > 0$, in the region $(m)$ given by (\ref{e4.30}), we have, on 
account of (\ref{e3.41}) and (\ref{e3.44}), 
\begin{eqnarray}\label{e4.39}
|I(\pm)|_{in(m)} & < &(-)^{n-m} [\phi_+(y_1) \pm \phi_-(y_1)]\cdot[\phi_-(y_1)
\phi_+(y_2)]\cdot[\phi_-(y_2)\phi_+(y_3)]\cdots \nonumber \\
& & [\phi_-(y_{m-1})\phi_+(y_m)]\cdot|\phi_-(y_{m})\phi_+(y_{m+1}) - 
\phi_+(y_{m})\phi_-(y_{m+1})|\cdot \nonumber \\
& & [\phi_+(y_{m+1})\phi_-(y_{m+2})]\cdots[\phi_+(y_n)\phi_-(y_{n+1})]
\phi_+(y_{n+1}).
\end{eqnarray}
For $0 < y_1 < \alpha$, we have $\phi_+(y_1)>\phi_-(y_1)>0$, 
\begin{eqnarray}
\phi_+(y_1)\phi_-(y_1) = O(\epsilon) \nonumber \\
\phi_-^2(y_1) = O(\epsilon) \nonumber,   
\end{eqnarray}
and therefore
\begin{eqnarray}
[\phi_+(y_1) \pm \phi_-(y_1)] \phi_-(y_1) = O(\epsilon). \nonumber 
\end{eqnarray}
Along the sequence on the right hand side of (\ref{e4.39}), we find
$$\phi_+(y_2)\phi_-(y_2) = O(\epsilon), \cdots$$ 
$$\phi_+(y_{m-1})\phi_-(y_{m-1}) = O(\epsilon),$$
and since both $\phi_+(y_{m})\phi_-(y_{m})$ and $\phi_+(y_{m-1})\phi_-(y_{m-1})$
are $O(\epsilon),$ 
\begin{eqnarray}
& & \phi_+(y_m)[\phi_-(y_m)\phi_+(y_{m+1}) - \phi_+(y_m)\phi_-(y_{m+1})]
\phi_+(y_{m+1}) \nonumber \\
& = &  \phi_+(y_m)\phi_-(y_{m})\phi_+^2(y_{m+1}) - \phi_+^2(y_{m})
\phi_-(y_{m+1})\phi_+(y_{m+1}) \nonumber \\
& = &  O(\epsilon). \nonumber 
\end{eqnarray}
Likewise,
$$\phi_-(y_{m+2})\phi_+(y_{m+2}) =  O(\epsilon), ~~~~\cdots $$ 
$$\phi_-(y_{n+1})\phi_+(y_{n+1}) =  O(\epsilon). $$
Thus, in the region $(m)$, $I(\pm) = O(\epsilon^n)$, so is $J_m(\pm)$; i.e., 
\begin{eqnarray}\label{e4.40}
J_m(\pm) = O(\epsilon^n).
\end{eqnarray}
Likewise, we can show $J_{n+1}(\pm)=O(\epsilon^n)$ and that leads to 
(\ref{e4.25}).~~~~~~~~~~~QED

Substituting (\ref{e4.15}) into (\ref{e4.17}), and setting the 
coefficients of different powers of $\epsilon$ separately to be 
zero, we have 
\begin{eqnarray}\label{e4.41}
1 + a_0\delta_0(e) = 0, 
\end{eqnarray}
\begin{eqnarray}\label{e4.42}
a_0\delta_1(e) + a_1\delta_0(e)^2 = 0, 
\end{eqnarray}
\begin{eqnarray}\label{e4.43}
a_0\delta_2(e) + 2a_1\delta_0(e)\delta_1(e) + a_2 \delta_0^3(e) = 0, 
\hspace{5mm} {\it etc.}.  
\end{eqnarray}
Likewise, 
\begin{eqnarray}\label{e4.44}
1 + b_0\delta_0(od) = 0, 
\end{eqnarray}
\begin{eqnarray}\label{e4.45}
b_0\delta_1(od) + b_1\delta_0(od)^2 = 0, 
\end{eqnarray}
\begin{eqnarray}\label{e4.46}
b_0\delta_2(od) + 2b_1\delta_0(od)\delta_1(od) + a_2 \delta_0^3(od) = 0, 
\hspace{5mm} {\it etc.}.  
\end{eqnarray}
(We recognize that, in view of (\ref{e4.19}) - (\ref{e4.24}), $a_n$ and
$b_n$ also contain $\epsilon$-dependence,, although each is $O(1)$ to the
leading order. We likewise permit the $\delta_n$ to have implicit 
$\epsilon$-dependences so that they may not be fixed uniquely by (\ref{e4.15}).
We have removed the ambiguity by requiring that when (\ref{e4.15}) is 
substituted into (\ref{e4.17}) - (\ref{e4.18}), the resulting equation 
should be true to each {\it explicit} power of $\epsilon$, 
disregarding 
the implicit entrance of $\epsilon$ through the $a_n$, $b_n$ and $\delta_n$.
The important point is that if $\Delta_e$, $\Delta_{od}$ are given by 
(\ref{e4.15}) with (\ref{e4.41}) - (\ref{e4.46}), then (\ref{e4.17}) - 
(\ref{e4.18}) will follow.)

To $O(\epsilon)$, 
\begin{eqnarray}\label{e4.47}
{\sf and}\hspace{10mm}\left.\begin{array} {ccccc}
2\Delta_e & = & \lambda\delta_0(e)& = & - \lambda/a_0\\
2\Delta_{od} & = & \lambda\delta_0(od) & = & - \lambda/b_0.  
\end{array}
\right. 
\end{eqnarray}
Since
\begin{eqnarray}\label{e4.48}
\frac{\int\limits_0^\infty \phi_-(y)\phi_+(y) dy}
{\int\limits_0^\infty \phi_+^2(y)dy} = O(\epsilon)
\end{eqnarray}
We find, from (\ref{e4.19}) - (\ref{e4.20}), 
\begin{eqnarray}\label{e4.49}
a_0 = \int\limits_0^\infty\phi_+^2(y)dy + O(\epsilon) \nonumber \\
b_0 = - \int\limits_0^\infty\phi_+^2(y)dy + O(\epsilon)
\end{eqnarray}
Thus 
\begin{eqnarray}\label{e4.50}
{\sf and} \hspace{15mm} \left.\begin{array}{ccc} 
\Delta_e & = & - [ \frac{\lambda}{2}/\int\limits_0^{\infty}\phi_+^2(y)dy] 
+ O(\epsilon^2) \\
\Delta_{od} & = & + [ \frac{\lambda}{2}/\int\limits_0^{\infty}\phi_+^2(y)dy] 
+ O(\epsilon^2) 
\end{array}
\right. 
\end{eqnarray}
which lead to, neglecting $O(\epsilon^2)$, 
\begin{eqnarray}
\frac{1}{2}(E_{even}+E_{odd}) = ga -\frac{1}{4a^2} -\frac{9}{2g(2a)^5} -
\frac{89}{2g^2(2a)^8} + O(\frac{1}{g^3a^{11}}), \nonumber \\
-\Delta_e = \Delta_{od} = \frac{\lambda}{2}/\int \limits_0^{\infty} 
\phi_+^2 dx, \nonumber \\
\lambda = 8ga^2e^{-\frac{4}{3}ga^3}[1 - \frac{3}{8ga^3} - 
\frac{53}{256(ga^3)^2} + O(\frac{1}{(ga^3)^3})], \nonumber
\end{eqnarray}
and $\phi_+=e^{-S(+)}$ given by (\ref{e2.6}) and (\ref{e2.11}).

By using the asymptotic expansion of $S(+)$ in $\phi_+=e^{-S(+)}$,
\begin{eqnarray}
\phi_-(x) = \lambda \phi_+(x)\int\limits_x^{\alpha}\frac{dy}{\phi_+^2(y)},
~~~~~~~~x>-\alpha \nonumber
\end{eqnarray}
and (\ref{e4.41}) - (\ref{e4.46}), we can evaluate the coefficients of 
$(ga^3)^{-m}e^{-\frac{4}{3}nga^3}$ in the double series expansion 
(\ref{e1.33}) - (\ref{e1.34}) for $E_{even}$ and $E_{odd}$ in terms of 
definite integrals. In particular, the corresponding asymptotic expansion 
(\ref{e2.12}) of $E$ is {\it automatically} agreeing with 
$\frac{1}{2}(E_{even}+E_{odd})$ to
$O(\epsilon)$. The underlying reason is: from (\ref{e3.13}), a variation of  
$\alpha \rightarrow \alpha + \delta \alpha$ gives $\delta \phi_-/\delta \alpha
\leq O(\epsilon)$, which leads to $\leq O(\epsilon^2)$ to $\Delta_e$ and 
$\Delta_{od}$

As remarked in Section 3, different choices of $\pm \alpha$ within the range 
(\ref{e3.5}) give different $E$ and $\phi_{\pm}$
within $O(\epsilon^2)$; however, they lead to the same $\psi_{even(odd)}$ and
$E_{even(odd)}$.

\newpage

\section*{\bf 5. Remarks and Other Examples }
\setcounter{section}{5}
\setcounter{equation}{0}

(1) Ansatz of trial wave functions.

In the notation of (\ref{e1.1}) - (\ref{e1.4}), let all the minima of the positive 
potential
\begin{eqnarray}\label{e5.1}
V(x) =g^2 v(x)
\end{eqnarray}
be at points
\begin{eqnarray}\label{e5.2}
l_1, l_2, \cdots,l_N;
\end{eqnarray}
i.e., 
\begin{eqnarray}\label{e5.3}
V(l_j) = 0,
\end{eqnarray}
with $j = 1, \cdots, N$.
For each $j$, there exists a solution of the Hamilton-Jacobi equation (\ref{e1.9}):
\begin{eqnarray}\label{e5.4}
\frac{1}{2}(\frac{dS_0}{dx})^2 - v(x) = 0+.
\end{eqnarray}
According to (\ref{e1.13}), this solution can be written as
\begin{eqnarray}\label{e5.5}
S_0(j) = \pm \int\limits_{l_j}^x[2v(y)+(0+)]^{\frac{1}{2}}dy,
\end{eqnarray}
with the $\pm$ sign chosen to make
\begin{eqnarray}\label{e5.6}
S_0(j) \geq 0 ~~~~~~~~~~~~~~~{\sf everywhere}.
\end{eqnarray}
These $N$
\begin{eqnarray}\label{e5.7}
\phi(j) = e^{-gS_0(j)}
\end{eqnarray}
functions are all linearly independent and can be used as a set of trial 
functions. In the neighborhood of $x=l_j$, the corresponding $\phi(j)$ describes 
the correct  harmonic oscillator ground state behavior. The value of $\phi(j)$ 
at a different minimum $x = l_i$ ($i \not=j$) gives the barrier penetration 
amplitude from $j$ to $i$, similar to the instanton description$^{[1,2]}$. 
Therefore, the set (\ref{e5.7}) forms a convenient ansatz of trial wave  
functions for the $N$ low-lying eigenstates
of the Hamiltonian (\ref{e1.2}). In the case when $V(x)$ is periodic, from this 
ansatz we can construct the lowest energy band structure of the problem.

(2) The systematic procedure developed in Section 4 can be extended to a much larger
class of potentials with degenerate minima. The limitation depends on the 
applicability of equations similar to (\ref{e4.7}) - (\ref{e4.8}). Here we give two examples that illustrate this limitation.

Consider first the special potential
\begin{eqnarray}\label{e5.8}
V(x) = u[-\delta(x-l) -\delta(x+l) +q\delta(x)],
\end{eqnarray}
where $l$, $u$ and $q$ are positive constants with 
\begin{eqnarray}\label{e5.9}
q  \leq 1,
\end{eqnarray}
so that
\begin{eqnarray}\label{e5.10}
\int\limits_{-\infty}^{\infty} V(x)dx  \leq -u.
\end{eqnarray}
The ground state of $H = -\frac{1}{2}(d^2/dx^2) +V(x)$ is 
\begin{eqnarray}\label{e5.11}
\psi_{even}(x) = \psi_{even}(-x) = \left\{\begin{array}{ll}
e^{-\kappa (x-l)},&~~~~~~~~~~~~x>l\\
(1-\frac{u}{\kappa})e^{-\kappa (x-l)}+\frac{u}{\kappa}
e^{\kappa(x-l)}, &~~~~~~~~~~~~0<x<l
\end{array}
\right.
\end{eqnarray}
with its eigenvalue
\begin{eqnarray}\label{e5.12}
E_{ev} = -\frac{1}{2}\kappa^2.
\end{eqnarray}
It can be readily shown that
\begin{eqnarray}\label{e5.13}
\kappa =  u [1+ \frac{\kappa-qu}{\kappa+qu}e^{-2\kappa l}].
\end{eqnarray}
Thus, for $q \leq 1$, $\kappa$ is $>u$, consistent with (\ref{e5.10}). If we neglect
the barrier penetration factor $e^{-2\kappa l}$ then $\kappa \cong u$, as would be
the case for a single $-u\delta (x-l)$ potential.

Introduce $\phi_+(x)$ for this problem as the solution of
\begin{eqnarray}\label{e5.14}
(H-E)\phi_+(x) = 0,
\end{eqnarray}
with $\phi_+(\infty) = 0$, $\phi_+(l) = 1$ and
\begin{eqnarray}\label{e5.15}
E = -\frac{1}{2}u^2.
\end{eqnarray}
We have 
\begin{eqnarray}\label{e5.16}
~~~\phi_+(x) = \left\{\begin{array}{ll}
e^{-u (x-l)},&~~~~~~~~~~~~x>l\\
e^{u (x-l)},&~~~~~~~~~~~~0<x<l\\
qe^{-u (x+l)} + (1-q)e^{u (x-l)},&~~~~~~~~~~~~-l<x<0\\
(q+2(1-q)e^{-2u l}) e^{u (x+l)}+(-1+q)e^{-2u l-u(x+l)},&~~~~~~~~~~~~x<-l.
\end{array}
\right.
\end{eqnarray}
Construct the Green's function $G_R$ in terms of $\phi_+(x)$ and 
$\phi_-(x)=\phi_+(-x)$:
\begin{eqnarray}\label{e5.17}
<x|G_R|y> = \frac{2}{\lambda}\cdot
\left\{\begin{array}{cc}
-\phi_+(x)\phi_-(y) + \phi_-(x)\phi_+(y) &,~~~ {\sf for}\hspace{2mm} x < y \\
0 &,~~~ {\sf for}\hspace{2mm} x > y 
\end{array}
\right. 
\end{eqnarray} 
where, similar to (\ref{e3.7})
\begin{eqnarray}\label{e5.18}
\lambda = \phi'_+(x)\phi_-(x) -\phi'_-(x)\phi_+(x) = 2u(1-q)e^{-2ul}.
\end{eqnarray}
Consequently, as in (\ref{e3.22})
\begin{eqnarray}\label{e5.19}
(H-E)<x|G_R|y> = -\delta(x-y).
\end{eqnarray}
Because $E_{ev} < E$, the integral
\begin{eqnarray}\label{e5.20}
\int \limits_{-\infty}^{\infty}<x|G_R|y> \psi_{even}(y)dy 
\end{eqnarray}
is convergent. Therefore, as in (\ref{e4.5}),
\begin{eqnarray}\label{e5.21}
(H-E)[\psi_{even} +\Delta_{e} \int \limits_{-\infty}^{\infty}
<x|G_R|y>\psi_{even}(y)dy] = 0,
\end{eqnarray}
where $\Delta_e = E_{ev} - E$. However, unlike (\ref{e4.7}) - (\ref{e4.8}), we find
that
\begin{eqnarray}\label{e5.22}
\chi_{even} \equiv \psi_{even} +\Delta_e \int \limits_{-\infty}^{\infty}
<x|G_R|y>\psi_{even}(y)dy 
\end{eqnarray}
satisfies
\begin{eqnarray}\label{e5.23}
\chi_{even} = 0.
\end{eqnarray}

For the quartic potential, instead of (\ref{e5.23}), we have (\ref{e4.6}), which can
be written as 
\begin{eqnarray}\label{e5.24}
\psi_{even}(x) = \phi_+(x) -\Delta_{e} \int \limits_{-\infty}^{\infty}
<x|G_R|y>\psi_{even}(y)dy 
\end{eqnarray}
in accordance with (\ref{e4.7}); furthermore
\begin{eqnarray}\label{e5.25}
\psi_{even}(x) \rightarrow\phi_+(x)~~~~~~~~~~~{\sf as}~~x \rightarrow \infty.
\end{eqnarray}
Recall that equations (\ref{e3.30}) - (\ref{e3.32}) are applicable for any $V(x)$.
Assume
\begin{eqnarray}\label{e5.26}
V(x) \rightarrow \frac{g^2}{2} x^n~,~~~~~~~~~{\sf as}~~~x \rightarrow \infty;
\end{eqnarray}
then instead of (\ref{e3.34}), we have for large $x$
\begin{eqnarray}\label{e5.27}
\theta = \frac{n}{4}\frac{1}{x} - \frac{E}{gx^{n/2}} + O(\frac{1}{x^2} or
\frac{1}{x^{2+\frac{n}{2}}}).
\end{eqnarray}
For $n>2$, $\theta$ is insensitive to $E$. The marginal case is $n=2$; in  
that case (\ref{e5.24}) remains valid, but both terms on its right hand side 
are important as $x \rightarrow \infty$. For the potential (\ref{e5.8}) of 
the above example, we have $n=0$; therefore (\ref{e5.24}) is not valid, 
being replaced by (\ref{e5.22}) - (\ref{e5.23}).

(3) We will now examine the marginal case $n=2$. Assume
\begin{eqnarray}\label{e5.28}
V(x) = g^2v(x) = \left\{\begin{array}{ccc}
\frac{1}{2}g^2(x - l)^2 & \hspace{5mm} {\sf for} \hspace{5mm} x > 0 \\
\Lambda \delta (x)\\
\frac{1}{2}g^2(x + l)^2 & \hspace{5mm} {\sf for} \hspace{5mm} x < 0. 
\end{array}
\right.
\end{eqnarray}
Let $\phi_+$ be the solution of
\begin{eqnarray}\label{e5.29}
H \phi_+ = -\frac{1}{2} \phi_+''(x) + V(x) \phi_+(x) = E \phi_+(x). 
\end{eqnarray}
Write
\begin{eqnarray}\label{e5.30}
\phi_+(x) = e^{-S(+)} 
\end{eqnarray}   
with $\phi_+(\infty) = 0$, and expand, as in (\ref{e1.6}) - (\ref{e1.7}),
\begin{eqnarray}\label{e5.31}
S(+) = g S_0(+) + S_1(+) + g^{-1}S_2(+) + \cdots 
\end{eqnarray}   
\begin{eqnarray}\label{e5.32}
E = g E_0 + E_1 + g^{-1}E_2 + \cdots 
\end{eqnarray}  
From (\ref{e5.29}), we have
\begin{eqnarray}\label{e5.33}
- S(+)'^2 + S(+)'' + 2g^2v(x) = 2E.     
\end{eqnarray}  
Substituting (\ref{e5.31}) and (\ref{e5.32}) into (\ref{e5.33}), and setting 
the coefficients of different powers of $g$ to be zero, we derive 
\begin{eqnarray}
S_0(+)' = (x - l), \nonumber
\end{eqnarray}  
which, together with the ``normalization'' condition $\phi_+(l) = 1$, gives
\begin{eqnarray}
S_0(+) = \frac{1}{2}(x - l)^2; \nonumber
\end{eqnarray} 
in addition,
\begin{eqnarray}
E_0 = \frac{1}{2},~~~E_1 = E_2 = \cdots = 0 \nonumber
\end{eqnarray} 
and
\begin{eqnarray}
S_1(+) = S_2(+) = \cdots = 0. \nonumber 
\end{eqnarray} 
Thus, 
\begin{eqnarray}\label{e5.34}
\phi_+(x) = e^{-\frac{1}{2}g(x - l)^2} \hspace{5mm} 
{\sf for} \hspace{5mm} x > 0.
\end{eqnarray} 

At $x = 0+$,
\begin{eqnarray}\label{e5.35}
{\sf and}\hspace{10mm}\left.\begin{array}{ccc}
\phi_+(0+) & = & e^{-\frac{1}{2}g l^2} \\
\phi'_+(0+) & = & g l e^{-\frac{1}{2}g l^2}.
\end{array}
\right.
\end{eqnarray} 
On account of the $\Lambda \delta(x)$ term in $V(x)$, at $x=0-$ we have
\begin{eqnarray}\label{e5.36}
{\sf and}\hspace{10mm}\left.\begin{array}{ccl}
\phi_+(0-) & = & e^{-\frac{1}{2}g l^2} \\
\phi'_+(0-) & = & -2\Lambda \phi_+(0) +\phi_+'(0+) = (-2\Lambda +gl)
e^{-\frac{1}{2}g l^2}.
\end{array}
\right.
\end{eqnarray} 

For $x<0$, $\phi_+(x)$ satisfies the second order differential equation 
(\ref{e5.29})
\begin{eqnarray}\label{e5.37}
-\frac{1}{2} \phi_+'' + \frac{1}{2}g^2 (x+l)^2 \phi_+ = \frac{1}{2}g \phi_+,
\end{eqnarray}
whose general solution is
\begin{eqnarray}\label{e5.38}
\phi_+(x) = a\chi (x) + be^{-\frac{1}{2}g(x + l)^2}
\end{eqnarray}
where $a$ and $b$ are constants and
\begin{eqnarray}\label{e5.39}
\chi (x) = e^{-\frac{g}{2}(x + l)^2}\int\limits_0^x e^{g(y + l)^2}dy.
\end{eqnarray}
Because of the boundary conditions (\ref{e5.36}),
\begin{eqnarray}\label{e5.40}
a = 2[g l -\Lambda] e^{-g l^2} \hspace{10mm}{\sf and} 
\hspace{10mm} b = 1.
\end{eqnarray}
Hence, for $x<0$
\begin{eqnarray}\label{e5.41}
\phi_+(x) = e^{-\frac{1}{2}g (x+l)^2}[1+a \int\limits_0^x e^{g (y + l)^2}dy].
\end{eqnarray} 
Introduce, as before,
\begin{eqnarray}\label{e5.42}
\phi_-(x) = \phi_+(-x).
\end{eqnarray} 
we have, for $x>0$,
\begin{eqnarray}\label{e5.43}
\phi_-(x) = e^{-\frac{1}{2}g (x-l)^2}[1-a \int\limits_0^x e^{g (y - l)^2}dy].
\end{eqnarray} 

When $\Lambda=gl$, we have $a=0$ and the exact groundstate wave function given by
\begin{eqnarray}\label{e5.44}
\hspace{20mm}\left\{\begin{array}{l}
e^{-\frac{1}{2}g(x-l)^2},~~~~~~~x>0 \\
e^{-\frac{1}{2}g(x+l)^2},~~~~~~~x<0;
\end{array}
\right.
\end{eqnarray}
the corresponding groundstate energy is
\begin{eqnarray}\label{e5.45}
\frac{1}{2}g.
\end{eqnarray} 
For $\Lambda\not= gl$, the Hamiltonian $H$ can be written as
\begin{eqnarray}\label{e5.46}
H=H_0+h,
\end{eqnarray}
where
\begin{eqnarray}\label{e5.47}
H_0=-\frac{1}{2}\frac{d^2}{dx^2}+\left\{\begin{array}{l}
\frac{1}{2}g^2(x-l)^2,~~~~~~~x>0 \\
gl\delta(x)\\
\frac{1}{2}g^2(x+l)^2,~~~~~~~x<0,
\end{array}
\right.
\end{eqnarray}
and
\begin{eqnarray}\label{e5.48}
h=(\Lambda-gl)\delta(x).
\end{eqnarray}
Regarding $h$ as the perturbation, to the first order in $h$, the change of
energy is given by the expectation value of $h$, with (\ref{e5.44}) as the
unperturbed wave function; i.e.,
\begin{eqnarray}\label{e5.49}
E_{even} \cong \frac{1}{2}g + \frac{(\Lambda-gl)e^{-gl^2}}
{2\int\limits_0^{\infty}e^{-g(x-l)^2 dx}}.
\end{eqnarray}

We now turn to an alternative derivation of $E_{even}$ by using (\ref{e4.7})
and (\ref{e4.9}). Let $\phi_+(x)$ and $\phi_-(x)$ be the same functions that
satisfy (\ref{e5.29})-(\ref{e5.43}). From (\ref{e5.34}) and (\ref{e5.43}),
it follows that for $x$ and $y$ both positive definite,
\begin{eqnarray}\label{e5.50}
\phi_+(x)\phi_-(y)-\phi_-(x)\phi_+(y) = a\phi_+(x)\phi_+(y)[f(x)-f(y)],
\end{eqnarray}
where
\begin{eqnarray}\label{e5.51}
f(x) = \int\limits_0^x e^{g(z-l)^2}dz,
\end{eqnarray}
and
\begin{eqnarray}\label{e5.52}
\phi'_+(x)\phi_-(y)-\phi'_-(x)\phi_+(y) = -g(x-l)[\phi_+(x)\phi_-(y)-
\phi_-(x)\phi_+(y)] +a\frac{\phi_+(y)}{\phi_+(x)}
\end{eqnarray}
which, at $x=y$, becomes
\begin{eqnarray}\label{e5.53}
\phi'_+(x)\phi_-(x)-\phi'_-(x)\phi_+(x) = a.
\end{eqnarray}
As in (\ref{e5.24}), we write
\begin{eqnarray}\label{e5.54}
\psi_{even}(x) = \phi_+(x) -\Delta_{e} \int \limits_{-\infty}^{\infty}
<x|G_R|y>\psi_{even}(y)dy, 
\end{eqnarray}
where
\begin{eqnarray}\label{e5.55}
E_{even} = \frac{g}{2} +\Delta_e
\end{eqnarray}
and 
\begin{eqnarray}\label{e5.56}
<x|G_R|y> = \left\{\begin{array}{ll}
\frac{2}{a}[\phi_-(x)\phi_+(y)-\phi_+(x)\phi_-(y)]~~~~~~~&,~~x<y \\
0~~~~~~~&,~~x>y.
\end{array}
\right.
\end{eqnarray}
For $y>x$, by using (\ref{e5.52}) - (\ref{e5.53}) we see that as
$x \rightarrow 0+$
\begin{eqnarray}\label{e5.57}
\frac{\partial}{\partial x}<x|G_R|y> \longrightarrow gl<0|G_R|y> -2[\phi_+(y)/\phi_+(0+)]. 
\end{eqnarray}
The function $\psi_{even}(x)$ and $\phi_+(x)$ are continuous at $x=0$,
but not their derivatives. We have
\begin{eqnarray}\label{e5.58}
\phi_+(0) = e^{-\frac{1}{2}gl^2}\\
\psi_{even}(0) = e^{-\frac{1}{2}gl^2} -\Delta_{e}
\int \limits_{-\infty}^{\infty}<0|G_R|y>\psi_{even}(y)dy\\ 
\phi'_+(0+) = gl e^{-\frac{1}{2}gl^2}
\end{eqnarray}
and
\begin{eqnarray}\label{e5.61}
\psi'_{even}(0+) = gl \psi_{even}(0) +2\Delta_{e} e^{\frac{1}{2}gl^2}
\int \limits_0^{\infty}\phi_+(x)\psi_{even}(x)dx.
\end{eqnarray}

From the Schroedinger equation with $V(x)$ given by (\ref{e5.28}), the
groundstate wave function must satisfy
\begin{eqnarray}\label{e5.62}
\frac{\psi'_{even}(0+)}{\psi_{even}(0)} = - \frac{\psi'_{even}(0-)}
{\psi_{even}(0)} = \Lambda.
\end{eqnarray}
Substituting (\ref{e5.61}) into (\ref{e5.62}), we derive
\begin{eqnarray}\label{e5.63}
\Lambda -gl = \frac{2\Delta_e}{\psi_{even}(0)} e^{\frac{1}{2}gl^2}
\int \limits_0^{\infty}\phi_+(x)\psi_{even}(x)dx. 
\end{eqnarray}

To derive $\Delta_e$ to $O(e^{-gl^2})$, we need only the zero$^{{\sf th}}$
order approximation in the wave function
\begin{eqnarray}\label{e5.64}
\psi_{even} \cong \phi_+
\end{eqnarray}
which gives
\begin{eqnarray}\label{e5.65}
\Delta_e = \frac{(\Lambda-gl)e^{-gl^2}}{2\int\limits_0^{\infty}
\phi_+^2(x) dx} + O(e^{-2gl^2})
\end{eqnarray}
in agreement with (\ref{e5.49}).

To derive the next order correction, we replace (\ref{e5.64}) by
\begin{eqnarray}\label{e5.66}
\psi_{even}(x) \cong \phi_+(x) -\Delta_{e} \int \limits_{-\infty}^{\infty}
<x|G_R|y>\phi_+(y)dy. 
\end{eqnarray}
From (\ref{e5.50}), (\ref{e5.53}) and (\ref{e5.56}), we have for $y>x>0$,
\begin{eqnarray}\label{e5.67}
<x|G_R|y> = 2\phi_+(x)\phi_+(y) \int \limits_x^y\phi_+^{-2}(z)dz.
\end{eqnarray}
Hence, the approximation (\ref{e5.66}) leads to
\begin{eqnarray}\label{e5.68}
\psi_{even}(0) \cong \phi_+(0) [1 - 2\Delta_e \int \limits_0^{\infty}
\phi_+^2(y)dy \int \limits_0^y \phi_+^{-2}(z)dz],
\end{eqnarray}
\begin{eqnarray}\label{e5.69}
\psi_{even}^{-1}(0) \cong \phi_+^{-1}(0) [1 + 2\Delta_e
\int \limits_0^{\infty} \phi_+^2(y)dy \int \limits_0^y \phi_+^{-2}(z)dz] 
\end{eqnarray}
and
\begin{eqnarray}\label{e5.70}
\int \limits_0^{\infty}\phi_+(x)\psi_{even}(x)dx & \cong &
\int \limits_0^{\infty}dx\phi_+(x)[\phi_+(x) -\Delta_{e}
\int \limits_x^{\infty}<x|G_R|y>\phi_+(y)dy] \nonumber \\
&& = \int \limits_0^{\infty}dy \phi_+^2(y)[1-2\Delta_e \frac
{\int \limits_0^{\infty} \phi_+^2(x)F(x)dx}
{\int \limits_0^{\infty} \phi_+^2(z)dz}],
\end{eqnarray}
where
\begin{eqnarray}\label{e5.71}
F(x) = \int \limits_x^{\infty} \phi_+^2(y)dy \int \limits_x^y
\phi_+^{-2}(z)dz.
\end{eqnarray}
Thus, (\ref{e5.69}) can also be written as
\begin{eqnarray}\label{e5.72}
\psi_{even}^{-1}(0) \cong \phi_+^{-1}(0) [1 + 2\Delta_e F(0)]
\end{eqnarray}
which, together with (\ref{e5.70}), gives
\begin{eqnarray}\label{e5.73}
\frac{1}{\psi_{even}(0)}
\int \limits_0^{\infty}\phi_+(x)\psi_{even}(x)dx \cong 
\frac{1}{\phi_+(0)}
\int \limits_0^{\infty}\phi_+^2(y)dy \nonumber \\
\cdot \{1- \frac{2\Delta_{e}}{\int \limits_0^{\infty}\phi_+^2(z)dz}
\int \limits_0^{\infty}\phi_+^2(x)[F(x)-F(0)]dx\}.
\end{eqnarray}
Because the potential $V(x) \rightarrow \frac{1}{2}g^2x^2$ as
$x \rightarrow \infty$, which is the marginal case $n=2$ in accordance with
(\ref{e5.27}), the corresponding integral $F(x)$ given by (\ref{e5.71}) has
a logarithmic divergence as the range of the $y$-integration
$\rightarrow \infty$. However, as we shall see, the difference $F(x)-F(0)$
and, therefore, $\Delta_e$ are well-defined.

We modify (\ref{e5.71}) by introducing a regulator $(1+\lambda y)^{-1}$ in
its integrand. Let
\begin{eqnarray}\label{e5.74}
F_{\lambda}(x) \equiv 
\int \limits_x^{\infty}(1+\lambda y)^{-1}\phi_+^2(y)dy
\int \limits_x^y\phi_+^{-2}(z)dz
\end{eqnarray}
and define
\begin{eqnarray}\label{e5.75}
F(x) - F(0) = \lim_{\lambda \rightarrow 0+}  [F_{\lambda}(x) - F_{\lambda}(0)].
\end{eqnarray}
In  (\ref{e5.74}), the $z$-integration is
\begin{eqnarray}\label{e5.76}
\int\limits_x^{y}\phi_+^{-2}(z)dz = \int\limits_x^y  e^{g(z-l)^2}dz
\nonumber\\
= \frac{e^{g(y-l)^2}}{2g(y-l)} 
- \frac{e^{g(x-l)^2}}{2g(x-l)} 
+\int\limits_x^y \frac{e^{g(z-l)^2}}{2g(z-l)^2}dz .
\end{eqnarray}
The subsequent $y$-integration for the first term  
on  the right-hand side of (\ref{e5.76}) leads to:
\begin{eqnarray}\label{e5.77}
\int\limits_x^{\infty}(1+\lambda y)^{-1}\phi_+^2(y)
\frac{e^{g(y-l)^2}}{2g(y-l)} dy \nonumber \\
=-\frac{1}{2g(1+\lambda l)}[ln\lambda + ln\frac{|l-x|}
{1+\lambda x} ].
\end{eqnarray}
In the difference $F_{\lambda}(x) - F_{\lambda}(0)$, the above  $ln\lambda$
term cancels. The limit $\lambda \rightarrow 0+$ gives a finite $F(x)-F(0)$,
therefore, a finite ratio
\begin{eqnarray}\label{e5.78}
\frac{1}{\psi_{even}(0)}
\int \limits_0^{\infty}\phi_+(x)\psi_{even}(x)dx
\end{eqnarray}
in accordance with (\ref{e5.73}). The substitution of (\ref{e5.73}) into
(\ref{e5.63}) leads to a well-defined quadratic equation for $\Delta_e$;
its solution determines $\Delta_e$ to the $O(e^{-2gl^2})$.

\section*{\bf Acknowledgement }

Thanks to RIKEN, Brookhaven National Laboratory and to U.S. Department of
Energy\footnote{No. DE-AC02-98CH10886} for providing the facilities essential 
for the completion of this work.

\newpage

\section*{\bf Appendix }
\setcounter{section}{9}
\setcounter{equation}{0}

In this Appendix, we discuss further the difference between a $\psi$-function 
and a $\phi$-function, introduced in Section 1 (between (\ref{e1.8}) and
(\ref{e1.9}). For clarity,
we return  to the quartic potential case in  which Hamiltonian $H$  is given
by (\ref{e1.27}) and  its even and odd groundstate wave functions are 
$\psi_{even}$  and $\psi_{odd}$. As  in (\ref{e1.29}),
\begin{eqnarray}\label{a1}
\psi_{\pm}(x) = \frac{1}{2}[\psi_{even}(x) \pm \psi_{odd}(x)].~~~~~~~~~~(A1)
\nonumber
\end{eqnarray}
Let $\phi_{\pm}(x)$ and $E$ satisfy (\ref{e3.1})  -  (\ref{e3.6}). At 
$x=\pm\infty$,  $\psi_{\pm}(x)$    are well-behaved, with 
\begin{eqnarray}\label{a2}
\psi_{\pm}(\infty) = \psi_{\pm}(-\infty) =0.~~~~~~~~~~~~~~~~~(A2)\nonumber
\end{eqnarray}
However,  since
\begin{eqnarray}\label{a3}
(H-E)\phi_{\pm}(x) = 0,~~~~~~~~~~~~~~~~~~~~~~~~~~~~~~~~~~(A3)\nonumber
\end{eqnarray}
with $E \not=$ an eigenvalue of $H$, only
\begin{eqnarray}\label{a4}
\phi_+(\infty) = \phi_-(-\infty) =0,~~~~~~~~~~~~~~~~~~~~(A4)\nonumber
\end{eqnarray}
but  $\phi_+(-\infty)$ and $\phi_-(\infty)$ are  divergent.

From (\ref{e4.1}) -  (\ref{e4.5}), we express (\ref{e4.6}), 
$\chi_j(x) \propto \phi_+(x)$, as 
\begin{eqnarray}\label{a5}
\chi_j(x) = c_j \phi_+(x), ~~~~~~~~~~~~~~~~~~~(A5)\nonumber
\end{eqnarray}
where $c_j$ are constants and, as before, $j=even(e)$ or $odd(od)$. Thus, 
(\ref{e4.7}) and (\ref{e4.8}) now  become  
\begin{eqnarray}\label{a6}
\psi_{even}(x) = c_e \phi_+(x) -\Delta_{e} \int \limits_{-\infty}^{\infty}
<x|G_R|y>\psi_{even}(y)dy ~~~~~~~(A6)\nonumber
\end{eqnarray}
and
\begin{eqnarray}\label{a7}
\psi_{odd}(x) = c_{od} \phi_+(x) -\Delta_{od} \int \limits_{-\infty}^{\infty}
<x|G_R|y>\psi_{odd}(y)dy. ~~~~~~~~~~~~~(A7)\nonumber
\end{eqnarray}
Through $x \rightarrow -x$, it follows that
\begin{eqnarray}\label{a8}
\psi_{even}(x) = c_e \phi_-(x) -\Delta_{e} \int \limits_{-\infty}^{\infty}
<x|G_L|y>\psi_{even}(y)dy ~~~~~~~~~~~~~~(A8)\nonumber
\end{eqnarray}
and
\begin{eqnarray}\label{a9}
\psi_{odd}(x) = -c_{od} \phi_-(x) -\Delta_{od} \int \limits_{-\infty}^{\infty}
<x|G_L|y>\psi_{odd}(y)dy. ~~~~~~~~~~~~~~~(A9)\nonumber
\end{eqnarray}
Combining (A1) with (A6) - (A7) and using the  matrix
notation of (\ref{e4.10}), we derive
\begin{eqnarray}\label{a10}
\psi_+ = c_+ \phi_+ -G_R [\Delta_{+}\psi_+ + \Delta_-\psi_-]~~~~~~~~(A10)
\nonumber
\end{eqnarray}
and
\begin{eqnarray}\label{a11}
\psi_- = c_- \phi_+ -G_R [\Delta_-\psi_+ + \Delta_+\psi_-]~~~~~~~~~~~(A11)
\nonumber
\end{eqnarray}
where
\begin{eqnarray}\label{a12}
&c_{\pm} = \frac{1}{2}(c_e\pm c_{od})\nonumber\\
{\sf and}~~~~~~~~~~~~~~~~&~~~~~~~~~~~~~~~~~~~~~~~~~~~~~~~~~~~~~~~~~(A12)
\nonumber\\
&\Delta_{\pm} = \frac{1}{2}(\Delta_e\pm \Delta_{od}).\nonumber
\end{eqnarray}
Likewise, from (A8) - (A9)
\begin{eqnarray}\label{a13}
\psi_+ = c_- \phi_- -G_L [\Delta_{+}\psi_+ + \Delta_-\psi_-]~~~~~~~~~(A13)
\nonumber
\end{eqnarray}
and
\begin{eqnarray}\label{a14}
\psi_- = c_+ \phi_- -G_L [\Delta_-\psi_+ + \Delta_+\psi_-].~~~~~~~~~~~~(A14)
\nonumber
\end{eqnarray}
On account of (\ref{e3.48}) - (\ref{e3.49}) and the corresponding  formulas,
replacing $G_R$ by $G_L$, we find 
\begin{eqnarray}\label{a15}
{\sf as} ~~x  \rightarrow \infty,~~~~~~
&\psi_+(x) \rightarrow c_+ \phi_+(x) \rightarrow 0~~~~~~~~~~~~~~~~~(A15)
\nonumber\\
&\psi_-(x) \rightarrow c_- \phi_+(x) \rightarrow 0~~~~~~~~~~~~~~~~~(A16)
\nonumber
\end{eqnarray}
and
\begin{eqnarray}\label{a17}
{\sf as} ~~x  \rightarrow -\infty,~~~~~~
&\psi_+(x) \rightarrow c_- \phi_-(x) \rightarrow 0~~~~~~~~~~~~~~~~~(A17)
\nonumber\\
&\psi_-(x) \rightarrow c_+ \phi_-(x) \rightarrow 0,~~~~~~~~~~~~~~~~(A18)
\nonumber
\end{eqnarray}
in agreement with (A2). As we shall see, it  is convenient to choose 
\begin{eqnarray}\label{a19}
c_+ = 1~~~~~~~{\sf and}~~~~~~c_-=\epsilon = e^{-\frac{4}{3}ga^3}.~~~~~~~(A19)
\nonumber
\end{eqnarray}
Thus,
\begin{eqnarray}\label{a20} 
\psi_+(x) \rightarrow
\left\{\begin{array}{ll}
\phi_+(x) ~~~~~~~~~~~~~~& {\sf as} ~~x \rightarrow \infty  \\
\epsilon \phi_-(x) ~~~~~~~~~~~~~~& {\sf as} ~~x \rightarrow -\infty  
\end{array}
\right.~~~~~~~~~~~~~~~~~~~~~~~~~~~(A20)\nonumber 
\end{eqnarray}
and
\begin{eqnarray}\label{a21} 
\psi_-(x) \rightarrow
\left\{\begin{array}{ll}
\epsilon \phi_+(x) ~~~~~~~~~~~~~~& {\sf as} ~~x \rightarrow \infty  \\
\phi_-(x) ~~~~~~~~~~~~~~& {\sf as} ~~x \rightarrow -\infty  .
\end{array}
\right.~~~~~~~~~~~~~~~~~~~~~~~~~~~(A21)\nonumber
\end{eqnarray}

We now examine the expressions 
\begin{eqnarray}\label{a22}
\psi_{\pm} = e^{-{\bf S}(\pm)},&\nonumber\\
\phi_{\pm} = e^{-S(\pm)}&~~~~~~~~~~~~~~~~~~~~~~~~~~~~~~~~~~(A22)\nonumber
\end{eqnarray}
and the asymptotic expansions (neglecting $O(\epsilon)$ corrections)
\begin{eqnarray}\label{a23}
{\bf S}(x) = g{\bf S}_0(x) + {\bf S}_1(x) + g^{-1}{\bf S}_2(x) + 
\cdots&\nonumber\\
S(x) = gS_0(x) + S_1(x) + g^{-1}S_2(x) + \cdots.&~~~~~~~~~~~~~~~~(A23)
\nonumber
\end{eqnarray}
From (A20) and (A21), we see that 
\begin{eqnarray}\label{a24} 
{\bf S}(+) \rightarrow
\left\{\begin{array}{ll}
S(+) ~~~~~~~~~~~~~~& {\sf as} ~~x \rightarrow \infty  \\
\frac{4}{3}ga^3+S(-) ~~~~~~~~~~~~~~& {\sf as} ~~x \rightarrow -\infty  .
\end{array}
\right.~~~~~~~~~~~~~~~~~~~(A24)\nonumber
\end{eqnarray}
and
\begin{eqnarray}\label{a25} 
{\bf S}(-) \rightarrow
\left\{\begin{array}{ll}
\frac{4}{3}ga^3+S(+) ~~~~~~~~~~~~~~& {\sf as} ~~x \rightarrow \infty  \\
S(-) ~~~~~~~~~~~~~~& {\sf as} ~~x \rightarrow -\infty  .
\end{array}
\right.~~~~~~~~~~~~~~~~~~(A25)\nonumber
\end{eqnarray}
Since, neglecting $O(\epsilon)$, $\{{\bf S}_n(\pm)\}$ and  $\{S_n(\pm)\}$
satisfy the same set of differential equations  we  have 
\begin{eqnarray}\label{a26} 
{\bf S}_n(+) = S_n(+) ~~~~~~~~~~~~~~
& {\sf for} ~~x >-a +O(\frac{1}{\sqrt{ga}}),~~~~~~~~~~(A26)\nonumber
\end{eqnarray}
and
\begin{eqnarray}\label{a27} 
{\bf S}_0(+) =  \frac{4}{3}a^3+ S_0(-) ~~~~~~~~~~~~~~
& {\sf for} ~~x <-a ~~~~~~~~~~~(A27)\nonumber\\
{\bf S}_m(+) = S_m(-) ~~~~~~~~~~~~~~
& {\sf for} ~~x <-a~~~ {\sf and}~~~ m\not=0.~~~~~~~~~~~~~~(A28)\nonumber
\end{eqnarray}
Likewise,
\begin{eqnarray}\label{a29} 
{\bf S}_n(-) = S_n(-) ~~~~~~~~~~~~~~
& {\sf for} ~~x <a -O(\frac{1}{\sqrt{ga}}),~~~~~~~~~~~~(A29)\nonumber
\end{eqnarray}
and
\begin{eqnarray}\label{a31} 
{\bf S}_0(-) =  \frac{4}{3}a^3+ S_0(+) ~~~~~~~~~~~~~~
& {\sf for} ~~x >a ~~~~~~~~~~~~~~(A30)\nonumber\\
{\bf S}_m(-) = S_m(+) ~~~~~~~~~~~~~~
& {\sf for} ~~x >a~~~ {\sf and}~~~ m\not=0.~~~~~~~~~~~~~~(A31)\nonumber
\end{eqnarray}

We note that ${\bf S}_0(\pm)$ and $S_0(\pm)$ are both solutions of 
\begin{eqnarray}\label{a32} 
[{\bf S}'_0(\pm)]^2 = 2v = (x^2-a^2)^2~~~~~~~~~~~~~~~~~~~~~~~~~(A32)\nonumber
\end{eqnarray}
and
\begin{eqnarray}\label{a33} 
[S'_0(\pm)]^2 = 2v = (x^2-a^2)^2.~~~~~~~~~~~~~~~~~~~~~~~~~(A33)\nonumber
\end{eqnarray}
In accordance with (\ref{e2.11}), at all $x$
\begin{eqnarray}\label{a34} 
&S_0(+) = \frac{1}{3}(x-a)^2(x+2a)\nonumber \\
{\sf and}~~~~~~~~~~~~~~~~&~~~~~~~~~~~~~~~~~~~~~~~~~~~~~~~~~~~~~~~~~~~~~(A34)
\nonumber\\
&S_0(-) = \frac{1}{3}(x+a)^2(-x+2a).\nonumber
\end{eqnarray}
However, from (A26) - (A27) and (A29) - (A30),
\begin{eqnarray}\label{a35} 
{\bf S}_0(+) =
\left\{\begin{array}{ll}
S_0(+)=\frac{1}{3}(x-a)^2(x+2a) ~~~~~~~~~~~~~~& {\sf for} ~~x >-a \\
\frac{4}{3}a^3+S_0(-)=\frac{4}{3}a^3+\frac{1}{3}(x+a)^2(-x+2a) 
~~~~~~~~~~~~~~& {\sf for} ~~x <-a.
\end{array}
\right.~~~~~~~~~~~~(A35)\nonumber
\end{eqnarray}
and
\begin{eqnarray}\label{a36} 
{\bf S}_0(-) =
\left\{\begin{array}{ll}
\frac{4}{3}a^3+S_0(+)=\frac{4}{3}a^3+\frac{1}{3}(x-a)^2(x+2a) 
~~~~~~~~~~~~~~& {\sf for} ~~x >a\\
S_0(-)=\frac{1}{3}(x+a)^2(-x+2a) ~~~~~~~~~~~~~~& {\sf for} ~~x <a,
\end{array}
\right.~~~~~~~~~~~(A36)\nonumber
\end{eqnarray}
which are solutions of the Hamilton-Jacobi equation
\begin{eqnarray}\label{a37} 
\frac{1}{2}(\frac{d{\bf S}_0(\pm)}{dx})^2 - v(x) = 0+~~~~~~~~~~~~~~(A37)
\nonumber
\end{eqnarray}
with  $0+$  as the total energy. By requiring ${\bf S}_0(\pm)=\infty$  at
$x=\pm\infty$ and ${\bf S}_0(\pm)=0$ at $x=\pm a$ respectively, we derive
(A35) - (A36) in the limit $0+  \rightarrow 0$.

\section*{\bf References }

1. A. A. Balavin, A. M. Polyakov, A.S. Schwartz and Yu S. Tyupkin, Phys. Lett. 
\underline{{\bf 59B}}, 85(1975)

2. G. 't Hooft, Phys. Rev. Lett. \underline{{\bf 37}}, 8(1979) 

}

\end{document}